\documentclass[aps,showpacs,preprint,floatfix,prd]{revtex4-1}
\usepackage{graphicx}
\usepackage{amsmath}
\usepackage{dcolumn}
\usepackage{bm}
\usepackage[colorlinks=true,linkcolor=blue,citecolor=blue]{hyperref}

\newcommand{\Tr}{\ensuremath{\mathop{\mathrm{Tr}}}}
\usepackage{float} 
\usepackage[dvipsnames]{xcolor}
\usepackage{ulem}

\begin{document}
\title{Lattice Calculation of Nucleon Isovector Axial Charge with Improved Currents}


\author{Jian Liang$^{1,}$\footnote[1]{jian.liang@uky.edu}, Yi-Bo Yang$^{1}$,  Keh-Fei Liu$^{1,}$\footnote[2]{liu@pa.uky.edu},\\ Andrei Alexandru$^{2}$\footnote[3]{aalexan@gwu.edu}, Terrence Draper$^{1}$ and Raza Sabbir Sufian$^{1}$
\begin{center}
\includegraphics[scale=0.12]{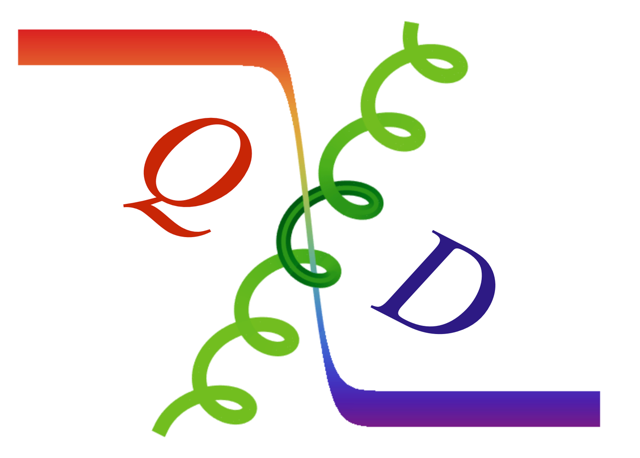}\\
($\chi$QCD Collaboration)
\end{center}
}
\affiliation{
$^{1}$\mbox{Department of Physics and Astronomy, University of Kentucky, Lexington, KY 40506, USA}
$^{2}$\mbox{Department of Physics, The George Washington University, Washington, DC 20052, USA}}

\begin{abstract}

We employ dimension-4 operators to improve the local vector and axial-vector currents
and calculate the nucleon isovector axial coupling $g^3_A$ with overlap valence on $2+1$-flavor domain wall fermion (DWF) sea.
Using the equality of $g^3_A$ from the spatial and temporal components of the axial-vector current as a normalization condition, 
we find that $g_A^3$ is increased by a few percent towards the experimental value.
The excited-state contamination has been taken into account with three time separations between the source and sink.
The improved axial charges $g_A^{3}(24I)=1.22(4)(3)$ and $g_A^{3}(32I)=1.21(3)(3)$ are obtained on a $24^3\times 64$ lattice at pion mass of 330 MeV
and a $32^3\times 64$ lattice at pion mass 300 MeV 
and are increased by 3.4\% and 1.7\% from their unimproved values, respectively. 
We have also used clover fermions on the same DWF configurations and find
the same behavior for the local axial charge as that with overlap fermions.

\end{abstract}

\maketitle

\section{introduction}
\label{introduction}

Lattice QCD, as a non-perturbative method for solving QCD problems,  
has achieved enormous success in the field of hadron physics.
However, there exist several quantities whose lattice results still {do not agree} with experimental values. 
One outstanding example is the nucleon isovector axial charge $g^3_A$.
The value of this charge, $1.2723(23)$~\cite{Agashe:2014kda}, has been well determined by the 
neutron $\beta$-decay experiments (see, for example,~\cite{Mendenhall:2012tz,Mund:2012fq}). 
However,
many lattice simulations (for recent results, please refer to~\cite{Alexandrou:2013joa,Bhattacharya:2013ehc,Owen:2012ts,   
Junnarkar:2014jxa,Syritsyn:2015nla,Ohta:2015aos,Yang:2015zja,Yoon:2016dij})
yield results as much as $\sim10\%$ lower than the experimental value. 
We note that some of $N_f=2$ lattice calculations do {have} results consistent with experiments~\cite{Horsley:2013ayv,Bali:2014nma,Abdel-Rehim:2015owa}, due to, for example,
the use of large source-sink time separations,
which is encouraging, but we think it is essential to have consistent $N_f = 2 +1$ results from different fermion actions at the physical pion mass 
with systematics such as the continuum and large volume extrapolations under control so that there is consensus from the community.
So it is still important to pursue lattice
studies on $g_A^3$ and nucleon structure in general both theoretically and technically to test the whole lattice methodology critically.

Furthermore, decomposing the nucleon spin into quark and gluon constituents,
i.e.  quark spin, quark orbital angular momentum, gluon spin and  gluon angular momentum, 
has long been an important issue.
The axial charge $g^q_A$ gives the intrinsic quark spin in the nucleon of flavor $q$.
So to solve the spin problem via lattice QCD,
an accurate $g^q_A$ from lattice must be demonstrated~\cite{Liu:2015nva}.
However, for the strange quark,
there have been a number of lattice calculations~\cite{QCDSF:2011aa,Engelhardt:2012gd,Babich:2010at,Abdel-Rehim:2013wlz,Alexandrou:2016mni} 
with their  $\Delta s=\langle N,i|\bar s \gamma_i\gamma_5 s |N,i\rangle$ ($i$ is the nucleon spin polarization direction) in the range from ${-}0.02$ to ${-}0.04$,
whose absolute values are several times smaller than those of the global fits of DIS experiments, e.g., 
$\Delta s\sim-0.11$~\cite{deFlorian:2009vb}, $\Delta s=-0.10(8) (x\in[10^{-3},1])$~\cite{Nocera:2014gqa} and $\Delta s\in [-0.11,-0.08]$~\cite{Adolph:2015saz},
leading to a large discrepancy.
So a lot of effort is still needed to close the gap and the accurate and precise calculation of $g^3_A$ can serve as a benchmark for all the lattice calculations.

The deviations between lattice results and the experimental values of $g^3_A$ can originate from lattice artifacts and lattice systematic errors,
such as the lack of chiral symmetry, finite volume effects, finite lattice spacing effects and excited-state contamination, 
if we believe that QCD is the correct description of the strong interaction.
One issue that escaped scrutiny is the fact that practically all the recent lattice calculations
use only local currents instead of improved currents or conserved currents for the chiral fermions such as domain wall fermion or overlap fermion.
With an $\mathcal{O}(a)$ improved fermion action, one needs to adopt the correspondingly $\mathcal{O}(a)$ improved current to remove the $\mathcal{O}(a)$ error.
And even with the conserved current, one needs $\mathcal{O}(a)$ improvement as well to obtain the improved conserved current
to make the matrix elements calculated to be $\mathcal{O}(a^2)$, especially for the off-forward case~\cite{Martinelli:1990ny}.
As we shall see from the present study,
we find that the two $g^3_A$ values obtained by inserting a point current $\bar\psi\gamma_5\gamma_{i}\psi$,$_{~i=1,2,3}$ or
$\bar\psi\gamma_5\gamma_4\psi$ are obviously different on each of the two lattice ensembles with $a = 0.1105(3)$ and $0.0828(3)$ fm respectively:
the one coming from $\bar\psi\gamma_5\gamma_4\psi$, denoted by $J_4^{A,P}$ (the superscript $P$ here stands for the point current),
is $10\%\sim20\%$ smaller than 
the one coming from $\bar\psi\gamma_5\gamma_{i}\psi$, which is denoted as $J_i^{A,P}$. 
Since the forward nucleon matrix element of $J_4^{A,P}$ must be calculated in a moving frame, and its signal-to-noise ratio (SNR) is not as good as the matrix element of $J_i^{A,P}$,
most previous lattice studies focus on $J_i^{A,P}$ only, so this may be the first observation of this kind of deviation.
This deviation, manifesting the asymmetry between spatial and temporal components on the lattice, is due to the finite lattice spacing artifact
and could be resolved by using the conserved or improved conserved lattice axial current. And hopefully, using the improved current may also improve the
final result at finite lattice spacing, leading to a more accurate result to be compared with the experimental value.

One can build a conserved vector current easily for Wilson-like fermion actions,
while constructing a conserved axial vector current is only viable for chiral fermions, such as the overlap fermion in our case~\cite{Hasenfratz:2002rp}.
Our future goal is to employ the conserved axial vector current (or its improved version) to calculate $g_A$ of the nucleon. 
In the meantime, as an exploratory study, we shall use dimension-4 currents in addition to the local ones in this work,
to see whether this kind of improvement can result in degenerate $\langle N|J_i^{A}|N'\rangle$ and $\langle N|J_4^{A}|N\rangle$ 
with common coefficients for different valence quark masses,
and lead to a $g^3_A$ value closer to experiments as well.

Since, in the isovector channel, the disconnected insertions are canceled between the
two degenerate light quark flavors,
we focus on the the connected insertion (CI) only in this work.
As a cautious test of our codes and results, we also carry out a test calculation using clover fermions for the valence quarks.

This paper is organized as follows.
In Sec.~(\ref{Numerical details}), we present the numerical details, including lattice setup, smeared-to-smeared 3-point 
function construction, and a new low-mode substitution scheme for the nucleon propagator.
Sec.~(\ref{improved currents}) introduces the improved currents used in this work.
Sec.~(\ref{Results})  contains the numerical results.
Then, we show the results of the test calculation on the clover fermions in Sec.~(\ref{a test of clover case}).
A  summary is given in Sec.~(\ref{summary}).

\section{Numerical details}
\label{Numerical details}
\subsection{Simulation Setup}

In this work, we use overlap fermions for the valence quark on $2+1$-flavor domain wall fermion (DWF) sea~\cite{Aoki:2010dy} to carry out the calculation.  
The effective quark propagator of the massive
overlap fermion is defined as the inverse of the operator $(D_c + m)$~\cite{Chiu:1998eu,Liu:2002qu}, where  $D_c$ is exactly chiral, i.e. $\{D_c, \gamma_5\} = 0$~\cite{Chiu:1998gp}, 
and can be expressed in terms of the overlap Dirac operator $D_{ov}$ as
\begin{eqnarray}
D_c=\frac{\rho D_{ov}}{1-D_{ov}/2} \textrm{ with }D_{ov}=1+\gamma_5\epsilon(\gamma_5D_{\rm w}(\rho)),
\end{eqnarray}
where $\epsilon$ is the matrix sign function and $D_{\rm w}$ is the Wilson Dirac operator with $\kappa$=0.2 (corresponding to parameter $\rho= 1.5$).
The RBC/UKQCD DWF gauge configurations used are from the $24^3\times64$ (24I) and $32^3\times64$ (32I) ensembles~\cite{Aoki:2010dy}.
The parameters of the ensembles are listed in Table~(\ref{table:r0}). 
We use 5 different quark masses with corresponding $m_{\pi}$ ranging from $\sim250$ MeV to $\sim400$ MeV on each of the two ensembles
to study the pion mass dependence of $g_A$ and $g_V$.
We compute 4 different source-sink separations, $8a$, $10a$, $11a$ and $12a$ on 24I and 
three separations $12a$, $14a$, $15a$ on 32I, 
to handle the excited-state contamination. 
The largest separations are $1.33$ fm for 24I and $1.24$ fm for 32I, respectively.

\begin{table}[!h]
\begin{center}
\begin{tabular}{cccccc}
Symbol & $L^3\times T$  &a (fm) & $m_l^{(s)}a$~~ &  {$m_{\pi}$} (MeV)  & $N_{cfg} $ \\
\hline
24I & $24^3\times 64$& ~0.1105(3) & 0.005   &330  & 203 \\
32I & $32^3\times 64$& ~0.0828(3) & 0.004   &300  & 309 \\
\hline
\end{tabular}
\end{center}
\caption{The parameters for the RBC/UKQCD configurations: label, spatial/temporal size, lattice spacing~\cite{Blum:2014tka}, 
 the degenerate light sea quark mass, the corresponding pion mass and the number of configurations used in this work.}
\label{table:r0}
\end{table}

To better control the SNR, 
we use two 12-12-12 (16-16-16 for 32I)
$Z_3$ noise grid sources with Gaussian smearing  at $t_{\rm{src}}=0$ and $t_{\rm{src}}=32$
and four 2-2-2 (1-1-1 for 32I) grid sources 
at $t_{\rm{sink}}$ which are 8, 10, 11 and 12 (12, 14 and 15 for 32I) time-slices away from the source positions with block smearing.
The notations such as 12-12-12 denote the intervals of the grid in the 3 spatial directions
(please see reference~\cite{Yang:2015zja} for more details).
Details regarding the simulation with overlap fermions are listed in Table~(\ref{valence_setup}).

\begin{table}[!h]
\begin{center}
\begin{tabular}{cccccccc}
Lattice & $\mathcal{G}_{\rm src}$ & $N_{\rm src}$ & $t_{\rm src}$ & $\mathcal{G}_{\rm sink}$ & $N_{\rm sink}$ & $(t_{\rm sink}-t_{\rm src})$ & $m_q^va$\\
\hline
      &                 &    &            &          & 5 & 0.88 fm  \\ \cline{6-7}
24I & 12-12-12 & 1 & (0, 32) & 2-2-2 & 3 & 1.11 fm  & (0.00809, 0.0102, 0.0135, 0.0160, 0.0203)\\ \cline{6-7} 
      &                 &    &            &          & 5 & 1.22 fm  \\ \cline{6-7}
      &                 &    &            &          & 5 & 1.33 fm  \\ \cline{6-7}
\hline
      &                 &    &            &          & 3 & 0.99 fm  \\ \cline{6-7}
32I & 16-16-16 & 1 & (0, 32) & 1-1-1 & 3 & 1.16 fm & (0.00585, 0.00765, 0.00885, 0.0112, 0.0152) \\ \cline{6-7}
      &                 &    &            &          & 3 & 1.24 fm  \\ \cline{6-7}
\hline
\end{tabular}
\end{center}
\caption{
The details of the overlap simulation in the valence sector, including the name of the lattice, the grid type of source $\mathcal{G}_{\rm src}$,
the number of source grids $N_{\rm src}$, the time positions of source $t_{\rm src}$, the grid type of sink $\mathcal{G}_{\rm sink}$,
the number of sink grids $N_{\rm sink}$, the source-sink separations $(t_{\rm sink}-t_{\rm src})$ and the bare valence quark masses $m_q^va$.}
\label{valence_setup}
\end{table}

The block smearing function can be defined as
\begin{equation}
\eta(x_0,x)=\sum_{|x'_i-{x_0}_i|\leq r,~i=1,2,3}\frac{1}{6}\left( P_{321}+P_{312}+P_{231}+P_{213}+P_{123}+P_{132} \right)\delta_{x',x},
\end{equation}
where $r$ is the smearing size, $P$ is a path of gauge links between spatial coordinate $x_0$ and $x'$, for instance, 
\begin{equation}
P_{321}=\tilde{U}_3\left({x'}_1,{x'}_2,{x'}_3; {x'}_1,{x'}_2,{x_0}_3\right)\tilde{U}_2\left({x'}_1,{x'}_2,{x_0}_3; {x'}_1,{x_0}_2,{x_0}_3\right)\tilde{U}_1\left({x'}_1,{x_0}_2,{x_0}_3; {x_0}_1,{x_0}_2,{x_0}_3\right),
\end{equation}
$x_i$ is the $i$ component of the coordinate $x$ and $U_i$ is a product of gauge links in the $i$ direction,
\begin{equation}
\tilde{U}_i\left(\vec{x}+N\hat{i},\vec{x}\right)=\prod_{n=0}^{N-1} U_i(\vec{x}+n\hat{i}), ~i=1,2,3.
\end{equation}
We average 6 types of paths to avoid bias on the order of the gauge links in the path.
Numerically, we have an algorithm to speed up the smearing process with such a definition, making the cost of the smearing proportional to $r$, rather than $r^3$~\cite{Yang:2016abc}.
The size of $r$ is tuned very close to the size of the Gaussian smearing at the source, so the source and sink 
are approximatively symmetric.
Different from the block case, the cost of Gaussian smearing is proportional to the iteration time~\cite{Gong:2013vja}, 
which is around 10 times slower than the block smearing for the smearing size used in this paper,
so we choose to employ the block smearing at the sink. 
More technical details regarding the calculation of the overlap operator, eigenmode deflation in the inversion of the fermion matrix, 
low-mode substitution (LMS) of random $Z_3$ grid source with mixed momenta, and the stochastic
sandwich method with LMS for constructing 3-point functions can be found in references~\cite{Li:2010pw,Gong:2013vja,Yang:2015zja}.
In the next subsection, we will discuss some improved numerical techniques of LMS.

\subsection{Stochastic sandwich method with smeared sink}

In our previous paper~\cite{Yang:2015zja}, we use point-sink when we calculate 3-point functions by means of the stochastic
sandwich method with LMS, but at the source position, we employ a Gaussian smeared source, leading to an asymmetry between {the} source
and sink. In that case, the excited-state contamination will be larger on the sink side and that asymmetry also causes difficulties
when we try to {do the fit}.

\begin{figure}[!h]
\includegraphics[width=0.49\textwidth]{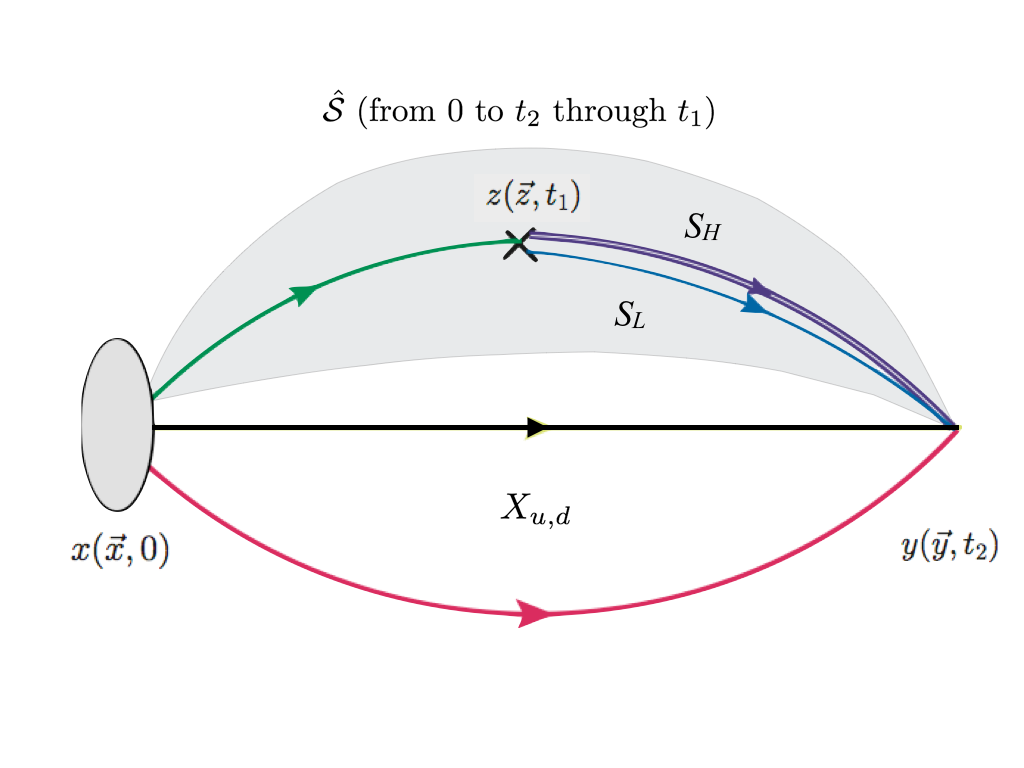}
\caption {The quark diagram of a nucleon correlator from position $x$ to $y$ with a connected insertion at $z$. 
The product of the quark propagators in the shadowed region constitutes the current-inserted propagator $\hat{\cal S}$. 
The propagator from the current position $z$ to the sink $y$ is decomposed into its low- and high-mode contributions ($S_L$ and $S_H$ respectively).
}
\label{fig:quark_diagram}
\end{figure}

The 3-point function of the nucleon can be expressed formally as~\cite{Yang:2015zja}, 
\begin{equation}
\label{C_3_1}
C_3(t_2,t_1)=\sum_{\vec{y}}\langle \Tr \left[\hat{\cal S}({\cal O},t_1;\vec{y},t_2;\vec{x},0) X_{u,d}(\vec{y},t_2;\vec{x},0;\Gamma,S',S'')\right] \rangle,
\end{equation}
where $\hat{\cal S}$ is a effective propagator coming from source $(\vec{x},0)$, going through the inserted current ${\cal O}(\vec{z},t_1)$, 
and ending at the sink $(\vec{y},t_2)$ (the shadowed region in Fig.~\ref{fig:quark_diagram});
$X_{u,d}$ is a functional of the other two propagators $S'$ and $S''$ (the black and red lines in Fig.~\ref{fig:quark_diagram}) without current insertion.
When performing LMS on the quark propagator between the sink and the current,
we decompose the sink part of $\hat{\cal S}$ into high-mode and low-mode parts,
\begin{equation}
\label{S_hat}
\hat{\cal S}({\cal O},t_1;\vec{y},t_2;\vec{x},0)=\sum_{\vec{z}} \left(S_L(\vec{y},t_2;\vec{z},t_1)+\sum_i\theta^i(\vec{y},t_2)\gamma_5\left(S^{i}_H(\vec{z},t_1;t_2)\right)^{\dagger}\gamma_5\right) {\cal O}(\vec{z},t_1)S(\vec{z},t_1;\vec{0},0),
\end{equation}
where $S_L(\vec{y},t_2;\vec{z},t_1)$ is the exact low-mode propagator propagating from $(\vec{z},t_1)$ to $(\vec{y},t_2)$,
$S^{i}_H(\vec{z},t_1;t_2)$ is the noise-estimated high-mode propagator propagating from $(t_2)$ to $(\vec{z},t_1)$,
$\theta^i(\vec{y},t_2)$ is the $Z_3$ noise vector helping to pick out the starting point $\vec{y}$ of $S^{i}_H$ and $i$ here is the index of noise vector.
If we focus on the sink spatial position $\vec{y}$ only, omit all other coordinate indices, and denote $\gamma_5\left(S^{i}_H(\vec{z},t_1;t_2)\right)^{\dagger}\gamma_5=\bar{S}^{i}_H$,
we can rewrite Eq.~(\ref{C_3_1}) and Eq.~(\ref{S_hat}) in a more compact form,
\begin{equation}
C_3=\sum_{\vec{y}}\langle \Tr \left[ \left(S_L(\vec{y})+\sum_i\theta^i(\vec{y})\bar{S}^{i}_H\right) {\cal O}S X_{u,d}(\vec{y};\Gamma,S',S'')\right] \rangle.
\end{equation}
The simple summation on $\vec{y}$ illustrates that it is a point-sink.
Since the low-mode propagator is an all-to-all propagator, we can pick out exactly any sink point $(\vec{y},t_2)$ without noise estimation.
This is why LMS can increase the SNR {greatly} especially for the low-mode dominated cases.

To implement the smeared-sink, we actually want 
\begin{equation}
\label{C_3_2_smear}
C_3=\sum_{\vec{y}}\langle\sum_{\vec{y}_1\vec{y}_2\vec{y}_3\in{\cal G}_{\rm{sink}}}\eta(\vec{y},\vec{y}_1)\eta(\vec{y},\vec{y}_2)\eta(\vec{y},\vec{y}_3) \Tr \left[ \left(S_L(\vec{y}_1)+\sum_i\theta^i(\vec{y}_1)\bar{S}^{i}_H\right) {\cal O}S X_{u,d}(\Gamma,S'(\vec{y}_2),S''(\vec{y}_3))\right] \rangle,
\end{equation}
where ${\cal G}_{\rm{sink}}$ denotes the sink grid and $\eta(\vec{x},\vec{y})$ is the smearing function from point $\vec{y}$ to $\vec{x}$, either Gaussian or block. 
In our practical calculations, we firstly apply smearing on the sink of $X_{u,d}$
\begin{equation}
X^S_{u,d}(\vec{y};\Gamma,S',S'')=\sum_{\vec{y}_2\vec{y}_3\in{\cal G}_{\rm{sink}}}\eta(\vec{y},\vec{y}_2)\eta(\vec{y},\vec{y}_3)X_{u,d}(\Gamma,S'(\vec{y}_2),S''(\vec{y}_3)),
\end{equation}
Eq.~(\ref{C_3_2_smear}) then becomes
\begin{equation}
\label{C_3_3_smear}
C_3=\sum_{\vec{y}}\langle\sum_{\vec{y}_1\in{\cal G}_{\rm{sink}}}\eta(\vec{y},\vec{y}_1)\Tr \left[ \left(S_L(\vec{y}_1)+\sum_i\theta^i(\vec{y}_1)\bar{S}^{i}_H\right) {\cal O}S X^S_{u,d}(\vec{y};\Gamma,S',S'')\right] \rangle.
\end{equation}
Note that the function $\eta$ is symmetric, i.e., $\eta(\vec{y},\vec{x})=\eta(\vec{x},\vec{y})$. So we can exchange the two summations and rewrite the above equation as
\begin{equation}
\label{C_3_4_smear}
C_3=\sum_{\vec{y}_1\in{\cal G}_{\rm{sink}}}\langle\Tr \left[ \left(S_L(\vec{y}_1)+\sum_i\theta^i(\vec{y}_1)\bar{S}^{i}_H\right) {\cal O}S \sum_{\vec{y}}\eta(\vec{y}_1,\vec{y}) X^S_{u,d}(\vec{y};\Gamma,S',S'')\right] \rangle.
\end{equation}
Here $\vec{y}_1\in{\cal G}_{\rm{sink}}$ is the point in the sink grid.
So for each $\vec{y}_1$, what we actually do {is} ``anti-smear" on  $X^S_{u,d}$, i.e., $\sum_{\vec{y}}\eta(\vec{y}_1,\vec{y}) X^S_{u,d}(\vec{y};\Gamma,S',S'')$
and then complete the final nucleon contraction.

It is well known that the smeared-to-smeared 2-point functions have poor SNR compared to that of the smeared-to-point correlators.
However for the smeared-to-smeared 3-point function to 2-point function ratios,
we find that the SNR is almost the same as the smeared-to-point ones, which should be attributed to the cancelations of the fluctuations when taking the ratio.
The fitting of the ratios are more stable and reliable as expected now for the symmetric source and sink.

\subsection{New LMS contraction scheme}

As shown above, the stochastic
sandwich method of 3-point function construction with LMS will eventually turn into
several 2-point function contractions of effective propagators.
In our previous implementation,
the nucleon correlation function with LMS is expressed as (copying
Eq.~(5) of reference~\cite{Yang:2015zja} here for convenience),
\begin{equation}
\label{nucleon_correlation_LMS}
\begin{split}
&C^{\rm{LMS}}\big(S_{NG},S_{NG},S_{NG}\big)\\
=&  C(S^H_{NG}, S^H_{NG}, S^H_{NG})\\
+&C\big(\sum_{x\in {\cal G}} \theta(x)S^L(x), S^H_{NG}, S^H_{NG}\big) + C\big(S^H_{NG}, \sum_{x\in {\cal G}} \theta(x)S^L(x), S^H_{NG}\big) + C\big(S^H_{NG}, S^H_{NG}, \sum_{x\in {\cal G}} \theta(x)S^L(x)\big)\\
+&\sum_{x\in {\cal G}} C\big(\theta(x) S^L(x), \theta(x) S^L(x), S^H_{NG}\big) + \sum_{x\in {\cal G}} C\big(\theta(x) S^L(x), S^H_{NG}, \theta(x) S^L(x)\big) +\\
+& \sum_{x\in {\cal G}} C\big(S^H_{NG}, \theta(x) S^L(x), \theta(x)S^L(x)\big)+\sum_{x\in {\cal G}} C\big(\theta(x) S^L(x), \theta(x) S^L(x), \theta(x) S^L(x)\big),
\end{split}
\end{equation}
where the functional $C$ means a normal contraction operation, $S^H_{NG}$ {denotes} the high-mode part of the noise grid propagator,
$S_L$ is the low-mode propagator, $\theta(x)$ is the $Z_3$ noise phase at the position $x$ which belongs to grid $\mathcal{G}$, and
$S_{NG}$ is the full noise grid propagator which is a combination of the high-mode part and the low-mode part.
The above equation is actually an expansion by dividing $S_{NG}$ into these two parts:
$S_{NG}=S^H_{NG}+\sum_{x\in\mathcal{G}}\theta(x)S_L(x)$.
In this old scheme, 
we have 4 functional $C's$ in line 2 and line 3 of Eq.~(\ref{nucleon_correlation_LMS}).
Since each one {entails} a normal contraction,
we need to do 4 times of nucleon contraction for these two lines.
In the last two lines of Eq.~(\ref{nucleon_correlation_LMS}),
functional $C's$ are within the summation of $x$,
and each summation consists of $N$ times of contraction operation,
where $N$ is the number of points in the grid ${\cal G}$,
so there are actually $4N$ times of nucleon contractions contained in these two lines.
Therefore, totally we need $4+4N$ times of normal contraction operations for a complete LMS contraction.

A better choice is
\begin{equation}
\label{nucleon_correlation_LMS2}
\begin{split}
&C^{\rm{LMS}}\big(S_{NG},S_{NG},S_{NG}\big)\\
=&\sum_{x\in {\cal G}} C\big(\theta(x) S^L(x)+S^H_{NG}, \theta(x) S^L(x)+S^H_{NG}, \theta(x) S^L(x)+S^H_{NG}\big)-(N-1)C(S^H_{NG}, S^H_{NG}, S^H_{NG}).
\end{split}
\end{equation}
Since the order of the grid summation $\sum_{x\in {\cal G}}$ and the $C$ functional can be {exchanged} optionally in the low-high-high case, i.e.,
$C\big(\sum_{x\in {\cal G}} \theta(x)S^L(x), S^H_{NG}, S^H_{NG}\big)=\sum_{x\in {\cal G}}C\big( \theta(x)S^L(x), S^H_{NG}, S^H_{NG}\big)$,
this expression is exactly equivalent to Eq.~(\ref{nucleon_correlation_LMS}),
\begin{equation}
\begin{split}
&\sum_{x\in {\cal G}} C\big(\theta(x) S^L(x)+S^H_{NG}, \theta(x) S^L(x)+S^H_{NG}, \theta(x) S^L(x)+S^H_{NG}\big)-(N-1)C(S^H_{NG}, S^H_{NG}, S^H_{NG})\\
=&  \sum_{x\in {\cal G}}C(S^H_{NG}, S^H_{NG}, S^H_{NG})-(N-1)C(S^H_{NG}, S^H_{NG}, S^H_{NG})\\
+&\sum_{x\in {\cal G}}C\big( \theta(x)S^L(x), S^H_{NG}, S^H_{NG}\big) + \sum_{x\in {\cal G}}C\big(S^H_{NG}, \theta(x)S^L(x), S^H_{NG}\big) + \sum_{x\in {\cal G}}C\big(S^H_{NG}, S^H_{NG}, \theta(x)S^L(x)\big)\\
+&\sum_{x\in {\cal G}} C\big(\theta(x) S^L(x), \theta(x) S^L(x), S^H_{NG}\big) + \sum_{x\in {\cal G}} C\big(\theta(x) S^L(x), S^H_{NG}, \theta(x) S^L(x)\big) +\\
+& \sum_{x\in {\cal G}} C\big(S^H_{NG}, \theta(x) S^L(x), \theta(x)S^L(x)\big)+\sum_{x\in {\cal G}} C\big(\theta(x) S^L(x), \theta(x) S^L(x), \theta(x) S^L(x)\big)\\
=&C^{LMS}\big(S_{NG},S_{NG},S_{NG}\big).
\end{split}
\end{equation}
However, Eq.~(\ref{nucleon_correlation_LMS2}) needs only $N+1$ times of normal contraction operation: $N$ times for the first term and $1$ time for the second one.
This new LMS contraction scheme reduces considerably the complication of coding, and takes only 1/4 of the computer time as compared to that in Eq.~(\ref{nucleon_correlation_LMS}).

\section{improved currents}
\label{improved currents}
Lattice QCD action is a discretized version of the continuum QCD action which is constructed on a hypercubic lattice.
Therefore, a simple local current, which is conserved in the continuum limit, is no longer conserved when the lattice spacing is finite.
For the vector case of Wilson-like fermions, the lattice version of conserved current is the point-split one~\cite{Doi:2009sq,Burger:2013jya}, 
which can be expanded in lattice spacing to have a local current and a derivative term in the next order in $a$:
\begin{equation}
\begin{split}
J_\mu^{ps}(x)=&\frac{1}{2}\left[\bar{\psi}(x)(\gamma_\mu-r)U_\mu(x)\psi(x+a\hat\mu)+\bar{\psi}(x+a\hat\mu)(\gamma_\mu+r)U^\dagger_\mu(x)\psi(x)\right]\\
=&\bar{\psi}(x)\gamma_\mu\psi(x)-a\bar\psi(x)\overleftrightarrow{D}_\mu\psi(x)+\mathcal{O}(a^2),
\end{split}
\end{equation}
where $\overleftrightarrow{D}=\frac{1}{2}(\overrightarrow{D}-\overleftarrow{D})$.
So to the lowest order, this is actually an $\mathcal{O}(a)$ modification to the point current.
On the other hand, even for lattice actions which are already improved to $\mathcal{O}(a^2)$, 
the matrix element of the local current as well as the point-split current contains $\mathcal{O}(a)$ errors. 
A fermion rotation defined in~\cite{Martinelli:1990ny,Heatlie:1990kg} is required to improve the currents
to $\mathcal{O}(a^2)$ {at} the same time
if one wants the discrete error of the final matrix element to be $\mathcal{O}(a^2)$,
leading to both the $\bar\psi(x)\overleftrightarrow{D}_\mu\psi(x)$ and $\partial_\nu(\bar\psi(x)\sigma_{\nu\mu}\psi(x))$ improvement terms in addition to the original point current.
For the overlap case, although it is not easy to {derive the explicit $\mathcal{O}(a)$ expansion} of the conserved current~\cite{Hasenfratz:2002rp},
one can follow reference~\cite{Glatzmaier:2014sya} to expand the $D_{\rm ov}=\frac{\rho}{a}(1-\frac{X}{\sqrt{X^\dagger X}})$ ($X$ is the Wilson kernel and $\rho$ is a parameter) 
order by order in the coupling constant $g_0$
by rewriting the square root term as an integral $\int\frac{d \sigma}{\pi}\frac{1}{\sigma^2+X^\dagger X}$ over a $\sigma$ parameter and 
Taylor expanding the resulting rational function as a series in the coupling constant, 
so besides the improvement current $\overleftrightarrow{D}_\mu$ coming from $X$, 
the improvement current containing $\sigma_{\mu\nu}$ 
also shows up in the coupling constant expansion of the square root in $D_{ov}$. 
Therefore, the general form of the improved vector current in our case can be expressed as
\begin{equation}
\begin{split}
J^V_{\mu}=&Z_V~\left(\bar{\psi}\gamma_\mu\hat\psi+f\bar{\psi}\overleftrightarrow{D}_\mu\hat\psi+g'\partial_\nu(\bar\psi\sigma_{\nu\mu}\hat\psi)\right).
\end{split}
\end{equation}
Whereas, for the axial vector case, only chiral fermions, like overlap~\cite{Hasenfratz:2002rp} and domain wall fermions~\cite{Boyle:2014hxa}, 
can have a chiral current which satisfies the anomalous Ward identity exactly and both are numerically expensive to calculate.
For the present work, we shall assume that the commonly used improvement terms $\bar\psi\gamma_5\sigma_{\mu\nu}\overleftrightarrow{D}_\nu\psi$ and 
$\partial_\mu(\bar\psi\gamma_5\psi)$ of axial current for general $\mathcal{O}(a)$ improved actions~\cite{Luscher:1996sc} apply to the overlap case
for the same reason explained in the vector case.
Therefore, the general form of improved axial current can be expressed as
\begin{equation}
\begin{split}
J^A_{\mu}=&Z_A~\left(\bar{\psi}\gamma_5\gamma_\mu\hat\psi+f'\partial_\mu(\bar{\psi}\gamma_5\hat\psi)+g\bar\psi\gamma_5\sigma_{\mu\nu}\overleftrightarrow{D}_\nu\hat\psi\right).
\end{split}
\end{equation}
$J^V_\mu$ and $J^A_\mu$ in the above two equations are the improved vector and axial vector currents;
$Z_V$ and $Z_A$ are the {normalization (finite renormalization)} constants;
$\hat\psi=(1 - 1/2 D_{ov}) \psi$ which {gives rise to the continuum-like effective quark propagator $(D_c + m)^{-1}$};
and $f$, $f'$, $g$ and $g'$ are the
coefficients of the improvement terms. We omit the lattice spacing $a$ in the above formulas for simplicity.
However, the $f'\partial_\mu(\bar\psi\gamma_5\hat\psi)$ term does not contribute to the forward matrix element when calculating $g_A$ and the
$g'\partial_\mu(\bar\psi\sigma_{\mu\nu}\hat\psi)$ term does not contribute to the unpolarized matrix element for $g_V$. 
And for chiral fermion in our case, we have $Z_A=Z_V\equiv Z$,
so the final improved currents used in this work are
\begin{equation}
\label{final_form}
\begin{split}
J^V_{\mu}=&Z~\left(\bar{\psi}\gamma_\mu\hat\psi+f\bar{\psi}\overleftrightarrow{D}_\mu\hat\psi\right),\\
J^A_{\mu}=&Z~\left(\bar{\psi}\gamma_5\gamma_\mu\hat\psi+g\bar\psi\gamma_5\sigma_{\mu\nu}\overleftrightarrow{D}_\nu\hat\psi\right).
\end{split}
\end{equation}

To sum up, the improvement operators are indeed inspired by the Wilson-like case~\cite{Martinelli:1990ny,Heatlie:1990kg}.
On the other hand, there are only two dimension-4 operators for the vector and axial currents. 
From the conserved vector current and chiral axial current for the overlap operator as formulated in reference~\cite{Hasenfratz:2002rp}, 
one can check the existences of these operators in the gauge coupling expansion~\cite{Glatzmaier:2014sya}. 
Thus, the coefficients $f$, $g$, $f'$, and $g'$ are functions of coupling $g_0$ and lattice spacing $a$, as is $Z_A$, and are determined separately for different lattices.
By using the equation of motion $\overrightarrow{D\!\!\!\!/}\psi=m\psi+\mathcal{O}(a)$,
the improvement current of the axial case can be written as
$a(\bar\psi\gamma_5\sigma_{\mu\nu}\overleftrightarrow{D}_\nu\hat\psi)=a(\mathcal{O}(a)+\partial_\mu\bar\psi\gamma_5\hat\psi+\mathcal{O}(a^2))$, 
where the lattice spacing $a$ is expressed
explicitly for clarity.
Since the $\partial_\mu \psi\gamma_5\hat\psi$ piece does not contribute to the forward matrix element, 
only the $a\mathcal{O}(a)$ term,
which is the difference between the continuum action and the lattice action,
survives to the lowest order of $a$ 
(the same argument can be found in, e.g.,~\cite{Bhattacharya:2005rb}),
such that our improvement current turns out to be an $\mathcal{O}(a^2)$ improvement to the local current
for the $g_A$ case with forward matrix elements.
Although it is commonly thought that the $\mathcal{O}(a^2)$ error should be small,
$g_A$ maybe one of the few exceptions.
We also find that the $\mathcal{O}(a^2)$ error can be as large as $20\%$ in our previous work of meson mass decomposition~\cite{Yang:2014xsa}.
In this work,
we will see that this $\mathcal{O}(a^2)$ improvement can indeed solve the discrepancy between the spatial and the temporal components of $g_A$.

The charges corresponding to the improved currents $J^V_{i}$ and $J^V_{4}$ are marked as $g_{V_i}$ and $g_{V_4}$; 
similarly, we also have the notations as $g_{A_i}$ and $g_{A_4}$.
Here $g_{V_i}$ and $g_{A_i}$ are averaged over values of $i=1,2,3$.
In the following, when using a latin letter, e.g.,  $i$, as the Dirac index, it ranges from $1$ to $3$, while greek letters range from $1$ to $4$.
Specifically, to calculate the connected 3-point functions of the improved currents, we need to carry out computation for each of the following currents
\begin{equation}
\begin{split}
J_i^{V,P}=&\bar\psi \gamma_i\hat\psi,~~~~~~
J_i^{V,D}=\bar\psi \overleftrightarrow{D}_i\hat\psi,\\
J_4^{V,P}=&\bar\psi \gamma_4\hat\psi,~~~~~~
J_4^{V,D}=\bar\psi \overleftrightarrow{D}_4\hat\psi,\\
J_i^{A,P}=&\bar\psi \gamma_5\gamma_i\hat\psi,~~~~
J_i^{A,D}=\bar\psi \gamma_5\sigma_{i\mu}\overleftrightarrow{D}_\mu\hat\psi,\\
J_4^{A,P}=&\bar\psi \gamma_5\gamma_4\hat\psi,~~~~
J_4^{A,D}=\bar\psi \gamma_5\sigma_{4i}\overleftrightarrow{D}_i\hat\psi.\\
\end{split}
\end{equation}
We use superscript $P$ or $D$ to denote the point currents and the dimension-4 currents with derivative, respectively.
We mark the corresponding charges of these currents as $g_{V_i}(P)$, $g_{V_{i}}(D)$, $g_{V_4}(P)$, $g_{V_{4}}(D)$, $g_{A_i}(P)$, $g_{A_{i}}(D)$,
$g_{A_4}(P)$ and $g_{A_{4}}(D)$ for further convenience.
Our goal is to calculate the improved isovector axial charge and to eliminate the deviation between the spatial and temporal parts,
that is to say, we will demand, after our improvement, $g_{A_i}=g_{A_4}$ and $g_{V_i}=g_{V_4}$ as our normalization conditions.
Using these two equations, we can solve for the coefficients $g$ and $f$ as
\begin{equation}
\label{g and f}
\begin{split}
f&=\frac{g_{V_4}(P)-g_{V_{i}}(P)}{g_{V_{i}}(D)-g_{V_{4}}(D)}\\
g&=\frac{g_{A_4}(P)-g_{A_{i}}(P)}{g_{A_{i}}(D)-g_{A_{4}}(D)}.
\end{split}
\end{equation}

The axial normalization constant $Z_A$ on the same lattices was determined in our previous work~\cite{Liu:2013yxz} through the chiral Ward identity for the pion.
In that case, none of the derivative terms contribute since the pion is at zero momentum;
the $Z_A$ is for the local current $\bar\psi\gamma_5\gamma_i\hat\psi$ only.
On the other hand, 
the two coefficients of the improvement currents can be determined non-perturbatively by Eq.~(\ref{g and f}) in this work.
So combining these two approaches we can compute all the factors appearing in Eq.~(\ref{final_form}).
$Z_Vg^3_V=1$ ($g^3_V$ is the improved vector charge) can be used as a further benchmark of the normalization constants
since one alway has $Z_V=Z_A$ for overlap fermions.

To calculate the above charges, we need to calculate the forward nucleon matrix element $\langle N(\vec{p},s)|\mathcal {O}|N(\vec{p},s')\rangle$.
This can be obtained via the 3-point function to 2-point function ratios $R_{C_3/C_2}$,
\begin{eqnarray}   \label{eq:ratio}
R_{C_3/C_2}(t_2,t_1)&=&\frac{\langle 0|  \Gamma_p{\hat{\chi}^S}(\vec{p},t_2){ \cal O}(t_1)\bar{\chi}^S(\vec{p},0)|0 \rangle}{\langle 0|\Gamma_e\hat{\chi}^S(\vec{p},t_2)\bar{\chi}^S(\vec{p},0)|0 \rangle},
\end{eqnarray}
where $\chi^S$ is the smeared proton interpolating field, and $\hat{\chi}^S$ is the same except for using $\hat{\psi}$ instead.
In the vector case,
$\Gamma_p=\Gamma_e\equiv\frac{1+\gamma_4}{2}$ is the non-polarized projector of the nucleon spin;
in the axial vector case, $\Gamma_p=\Gamma_i\equiv\frac{1+\gamma_4}{2}\gamma_5\gamma_i$ is the polarized projector.
When $t_2$ is large enough, there should exist a plateau,
which is denoted as $\bar{R}_{C_3/C_2}=R_{C_3/C_2}(t_2\to\infty,t_1\gg0)$.
$\bar{R}_{C_3/C_2}$ is a product of the desired matrix element $\langle N|{\cal{O}}|N\rangle$ and a kinematic factor $F_k$.
To extract the matrix elements, we need to compute these factors first.
For example if ${\cal O}=\bar\psi \Gamma\psi$, $\Gamma$ is some gamma matrix,
\begin{equation}
\begin{split}
F_k=\frac{\textrm{Tr}\left[\Gamma_p\frac{-ip\!\!\!/+m}{2E}\Gamma\frac{-ip\!\!\!/+m}{2E}\right]}{\textrm{Tr}\left[ \Gamma_e\frac{-ip\!\!\!/+m}{2E} \right]},
\end{split}
\end{equation}
and $m$ and $E$ are the nucleon mass and energy.
The matrix element can then be expressed as
$\langle N|{\cal{O}}|N\rangle=\bar{R}_{C_3/C_2}/F_k$.

\begin{table}[!h]
\begin{center}
{\begin{tabular}{cccc}
current & $F_k$ & $\vec{p}=\vec{0}$ & $\vec{p}^2=p_i^2$ \\
\hline
$J_i^{V,P}$ & $-i\frac{p_i}{E}$ & 0 &  $-i\frac{p_i}{E}$ \\
$J_4^{V,P}$ & $1$ & $1$ & 1 \\
$J_i^{V,D}$ & $i\frac{mp_i}{E}$ & 0 & $i\frac{mp_i}{E}$ \\
$J_4^{V,D}$ & $-m$ & $-m$ & $-m$ \\
\hline
$J_i^{A,P}$ & $-\frac{m^2+mE+p_i^2}{E(E+m)}$ & $-1$ & $-1$ \\
$J_4^{A,P}$ & $-i\frac{p_i}{E}$ & 0 & $-i\frac{p_i}{E}$ \\
$J_i^{A,D}$ & $\frac{m^2(m+E)+mp_i^2}{E(E+m)}$ & $m$ & $m$\\
$J_4^{A,D}$ & $i\frac{mp_i}{E}$ & 0 & $i\frac{mp_i}{E}$\\
\hline
\end{tabular}}
\end{center}
\label{K_factors}
\caption{Kinematic factors in the 3-point function to 2-point function ratios. The last two columns show the results in two special cases, which are used in this work.}
\end{table}

All the factors are listed in Table~(\ref{K_factors}). 
For the axial vector case, we choose the polarization index the same as the Dirac index $i$ for currents $J_i^{A,P}$ and $J_i^{A,D}$.
The index $i$ of the $p_i$ dependence for currents $J_4^{A,P}$ and $J_4^{A,D}$ comes from the polarized projector.
It is found that all improvement currents have the same structures as the local ones,
and the kinematic factors of the improvement currents just have one additional ``$-m$" multiplied to the factors of the corresponding local currents.
For $J_i^{V,P}$ and $J_i^{V,D}$, we need to carry out the calculations in a moving frame because the factors are proportional to the
nucleon momentum. For $J_4^{V,P}$ and $J_4^{V,D}$, we can do the calculation in the rest frame of the nucleon. This is the
reason why they are separated into two parts.
For the axial case, it is the other way around.

\section{Results}
\label{Results}

\subsection{Vector Case}

\begin{figure}[!h]
\includegraphics[width=0.49\textwidth,page=1]{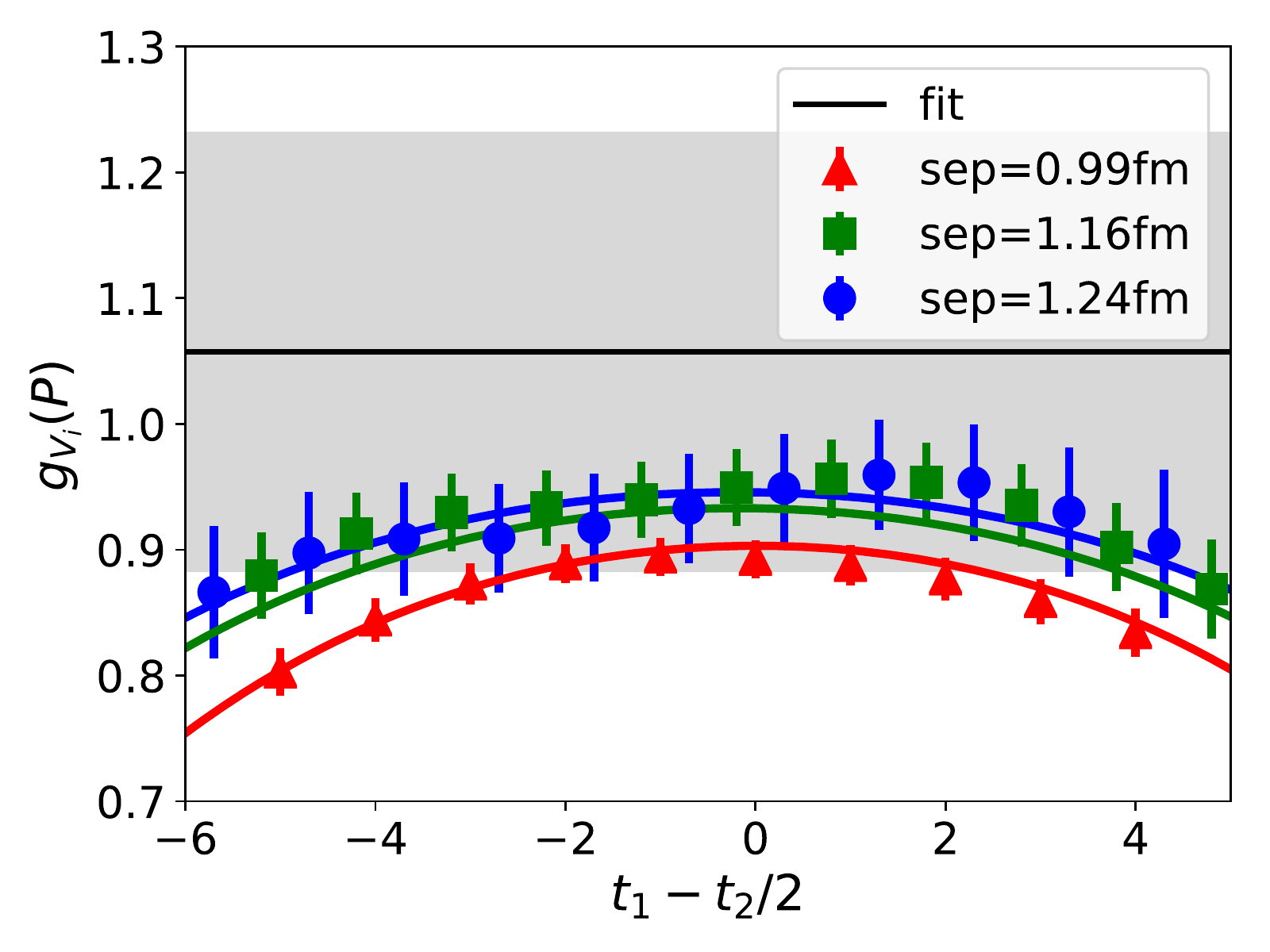}
\includegraphics[width=0.49\textwidth,page=1]{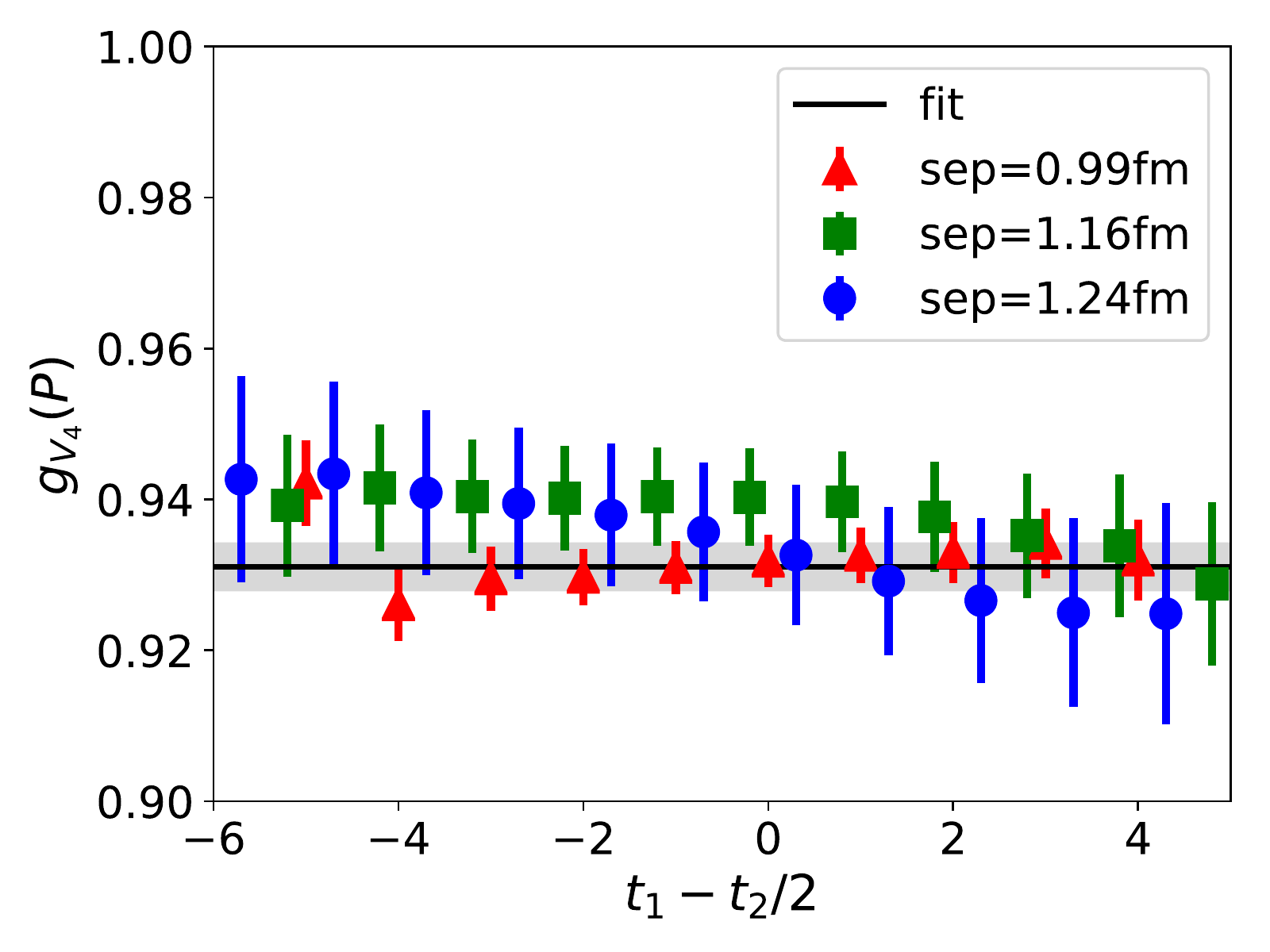}
\includegraphics[width=0.49\textwidth,page=1]{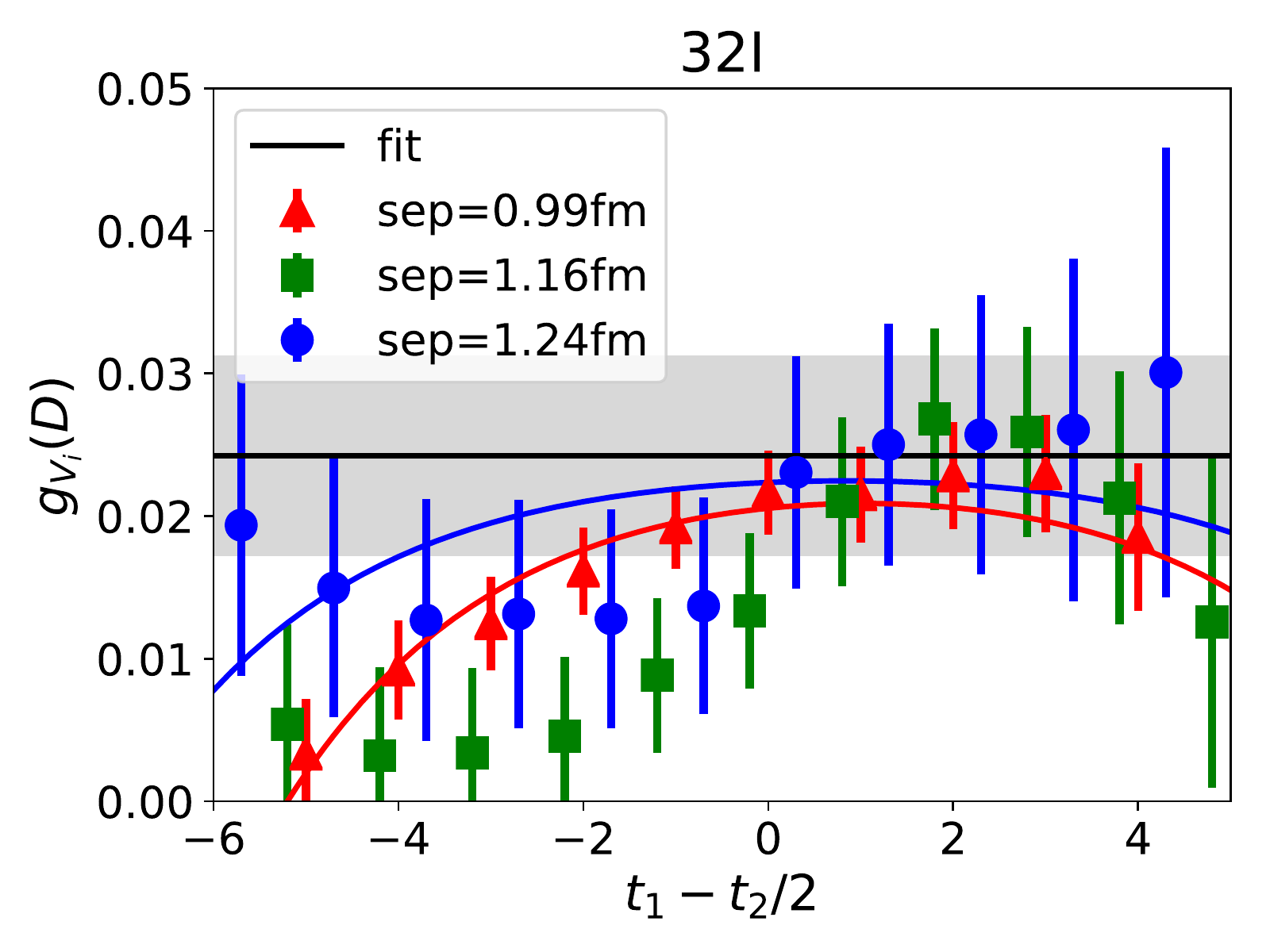}
\includegraphics[width=0.49\textwidth,page=1]{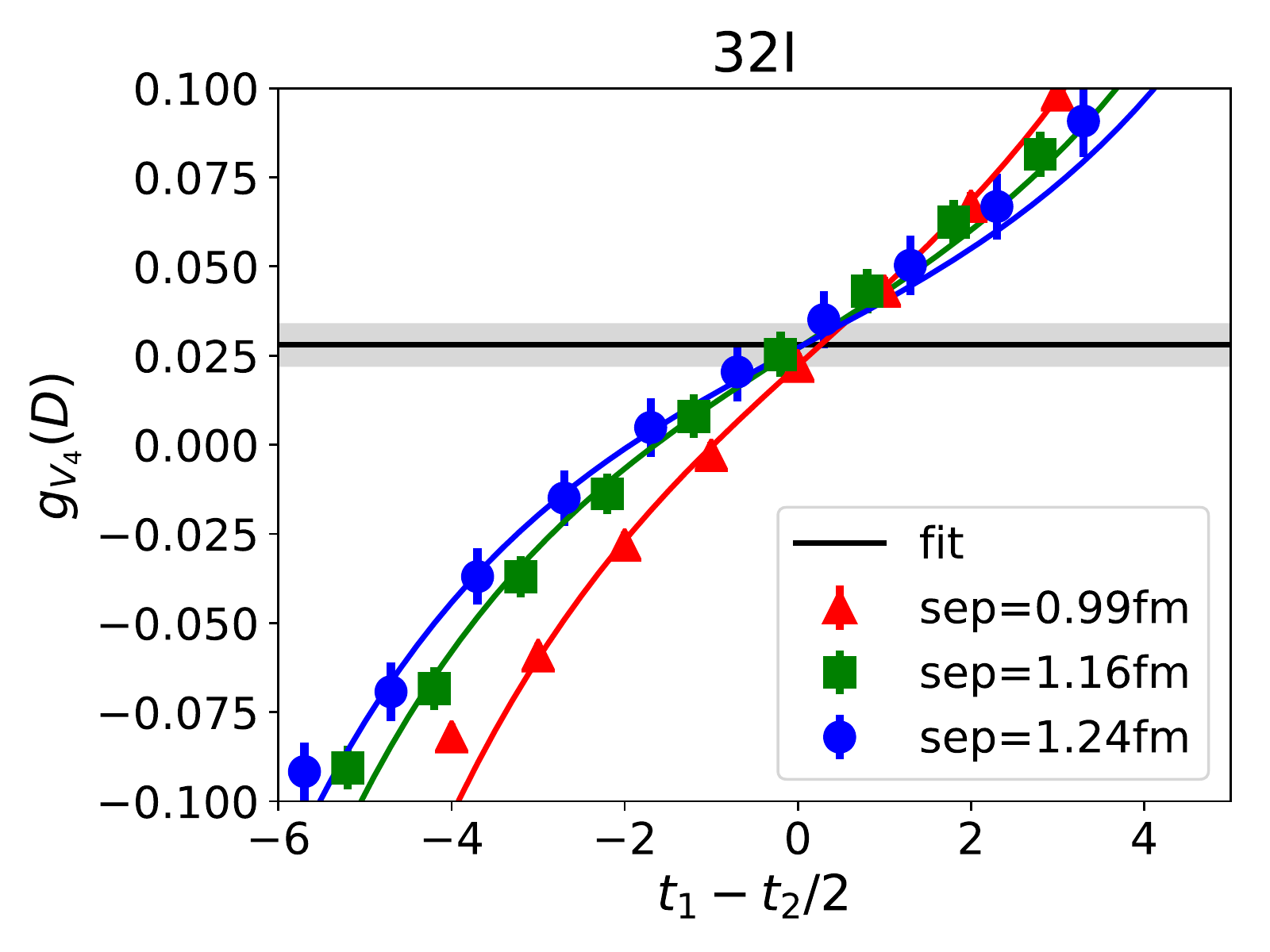}
\caption{Example of 2-state fits in the vector case on the 32I lattice at the unitary point. 
Points with error bars
are from lattice results and the curves are from the 2-state fits. For $g_{V_4}(P)$, it is too flat to use 2-state fit,
so we use a constant fit instead. The black lines are the final fit values and the gray bands indicate the fit errors. 
All the values are not renormalized.}
\label{32I_gV_m3}
\end{figure}

As we mentioned above in section~(\ref{improved currents}), 
we will compute $g_V$ to check whether the same $Z_A$ determined from the pion also applies to the proton case.
We use a 2-state fit to handle the ratios
\begin{equation}
R(t_1,t_2)=C_0+C_1e^{-\delta m (t_2-t_1)}+C_2e^{-\delta m t_1}+C_3e^{-\delta m t_2},
\end{equation}
where $\delta m$ is the energy difference between the first excited-state and the ground state,
$t_2$ is the source-sink separation and $t_1$ is the time slice with current insertion.
Constant $C_0$ is the desired matrix element, coefficients $C_1$ and $C_2$ are related to the transition between the ground state and the first excited state, 
while $C_3$ accounts for the excited-state to excited-state contribution.
So in different channels their significance can be different; we need to pick out the significant terms in order to get a stable fit. 
The $C_3$ term is alway nonsignificant for the vector case and is removed from the 2-state fit.
The difference caused by adding different terms in the fit will be considered as a systematic uncertainty.
As an example, the fitted results on 32I at the unitary point can be found in Fig.~\ref{32I_gV_m3}.
For $g_{V_4}(P)$, it is too flat to apply the 2-state fit,
so we use a constant fit instead.
For $g_{V_{4}}(D)$, there is no plateau on the plot, which is presumably due to the fact that we used $\overrightarrow{D}$ rather than $\overleftrightarrow{D}$
in our 3-point function contraction code.
However, a 2-state fit can handle this case very well, as the lattice data points almost all lay on the fit curves.
(We also tested $\overleftrightarrow{D}$ in the clover case using sink-sequential methods and
the final results are not affected; the figures will be shown in the next section.)
For $g_{V_i}(P)$, the final error band is much larger than the error of the lattice data,
which is because the fitted $\delta m$ is small and the excited-state contributions cannot be accurately fixed by the data.
We also tried to use a constant fit for $g_{V_i}(P)$ and the results are consistent with the 2-state fit, 
but the $\chi^2/d.o.f.$ is around 2 which is unacceptable.
For $g_{V_{i}}(D)$, since it is noisier than the other 3 cases, a 2-state fit gives a result consistent with the constant fit;
we choose to use the 2-state fit as shown in Fig.~\ref{32I_gV_m3}.

\begin{figure}[!h]
\includegraphics[width=0.49\textwidth,page=1]{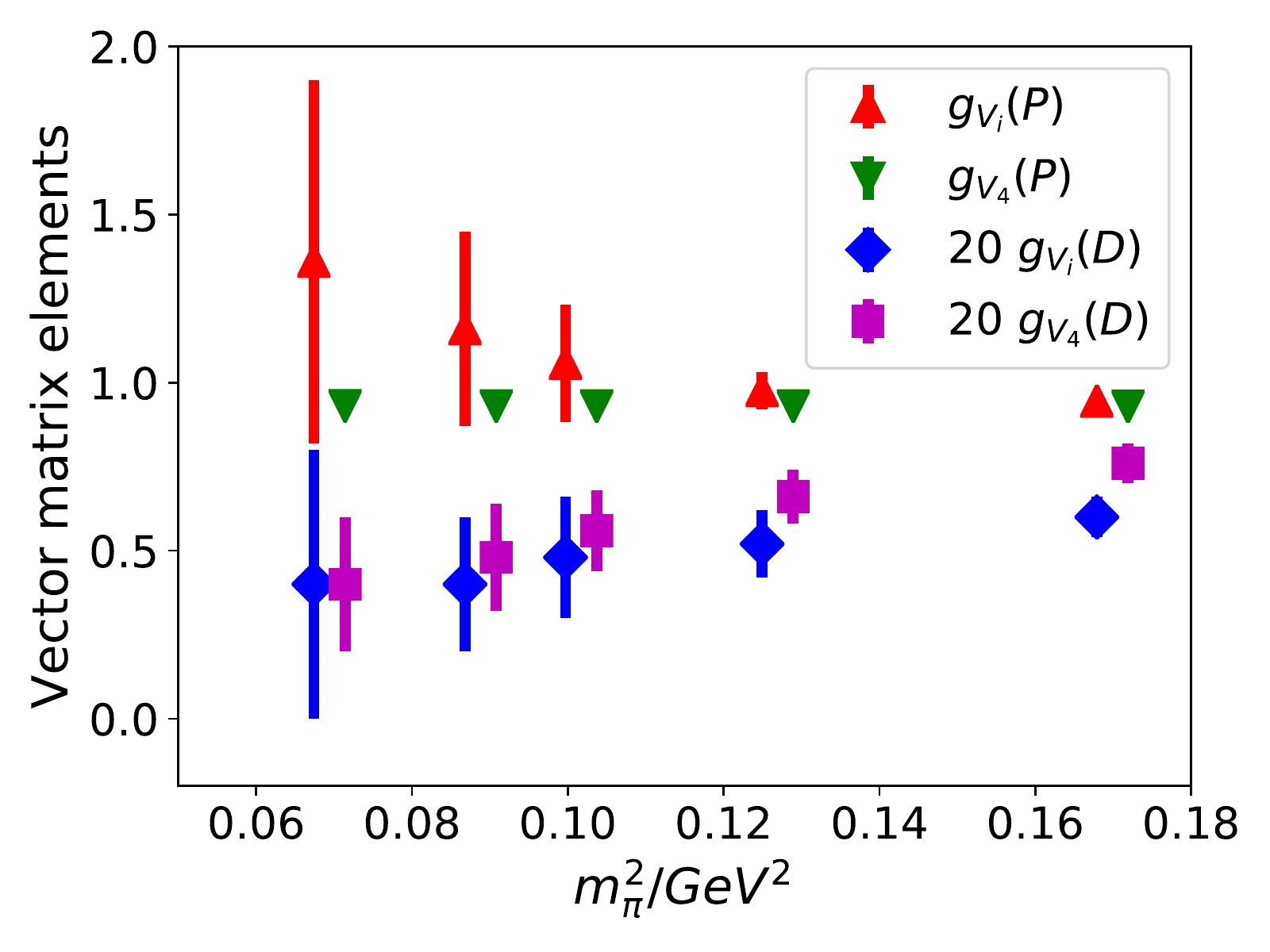}
\caption{Bare results of the vector cases as a function of pion mass squared for 32I. $g_{V_{i}}(D)$ and $g_{V_{4}}(D)$ are rescaled by a factor of 20 for clarity.}
\label{32I_gV}
\end{figure}

The pion mass dependence is shown in Fig.~\ref{32I_gV}. $g_{V_4}(P)$ keeps constant when $m_\pi$ changes and
$g_{V_i}(P)$ at each $m_{\pi}$ is consistent with $g_{V_4}(P)$ within errors;
$g_{V_{i}}(D)$ decreases a little as the pion mass decreases but the values are all consistent with $g_{V_4}(D)$.
Given this situation, the equation $g_{V_{i}}=g_{V_{4}}$ is already satisfied within errors so we cannot determine a unique factor $f$ for the improvement term.
In other words, there is no obvious need for this kind of improvement in the vector channel, since no evident deviation between the
spatial and temporal parts is observed.

The results from 24I are similar. 
Using the bare value of $g_{V_4}(P)$,
we find that $Z_Vg_{V_4}(P)=1$ within errors with $Z_V=Z_A$ determined from pion using Ward identity~\cite{Liu:2013yxz},
which means the pion $Z_A$ also applies to the proton case.
So we will use the normalization constant provided in the above reference,
which is, $Z_A=1.111(6)$ for 24I and $Z_A=1.086(2)$ for 32I.

\subsection{Axial Vector Case}

\begin{figure}[!h]
\includegraphics[width=0.49\textwidth,page=1]{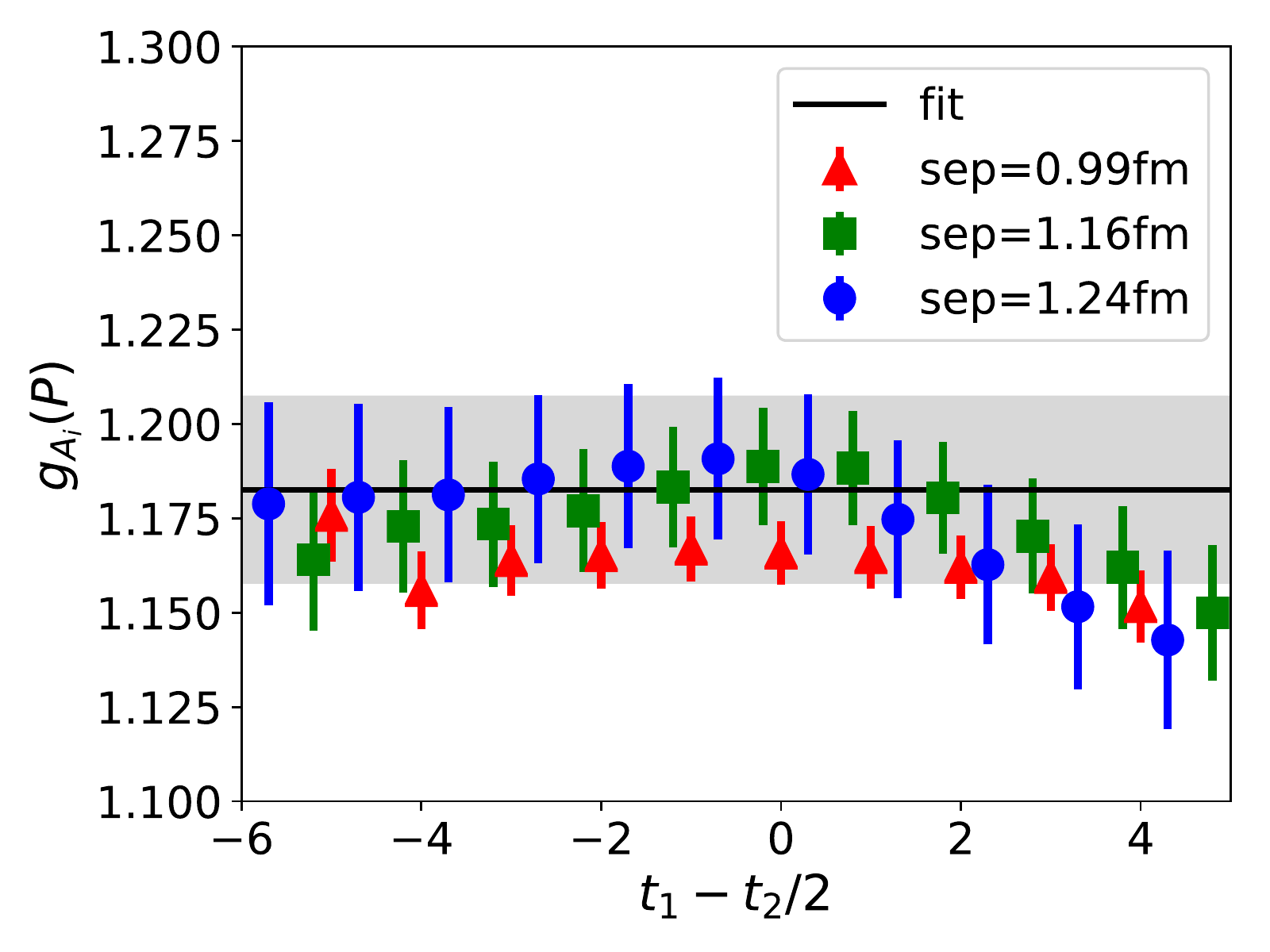}
\includegraphics[width=0.49\textwidth,page=1]{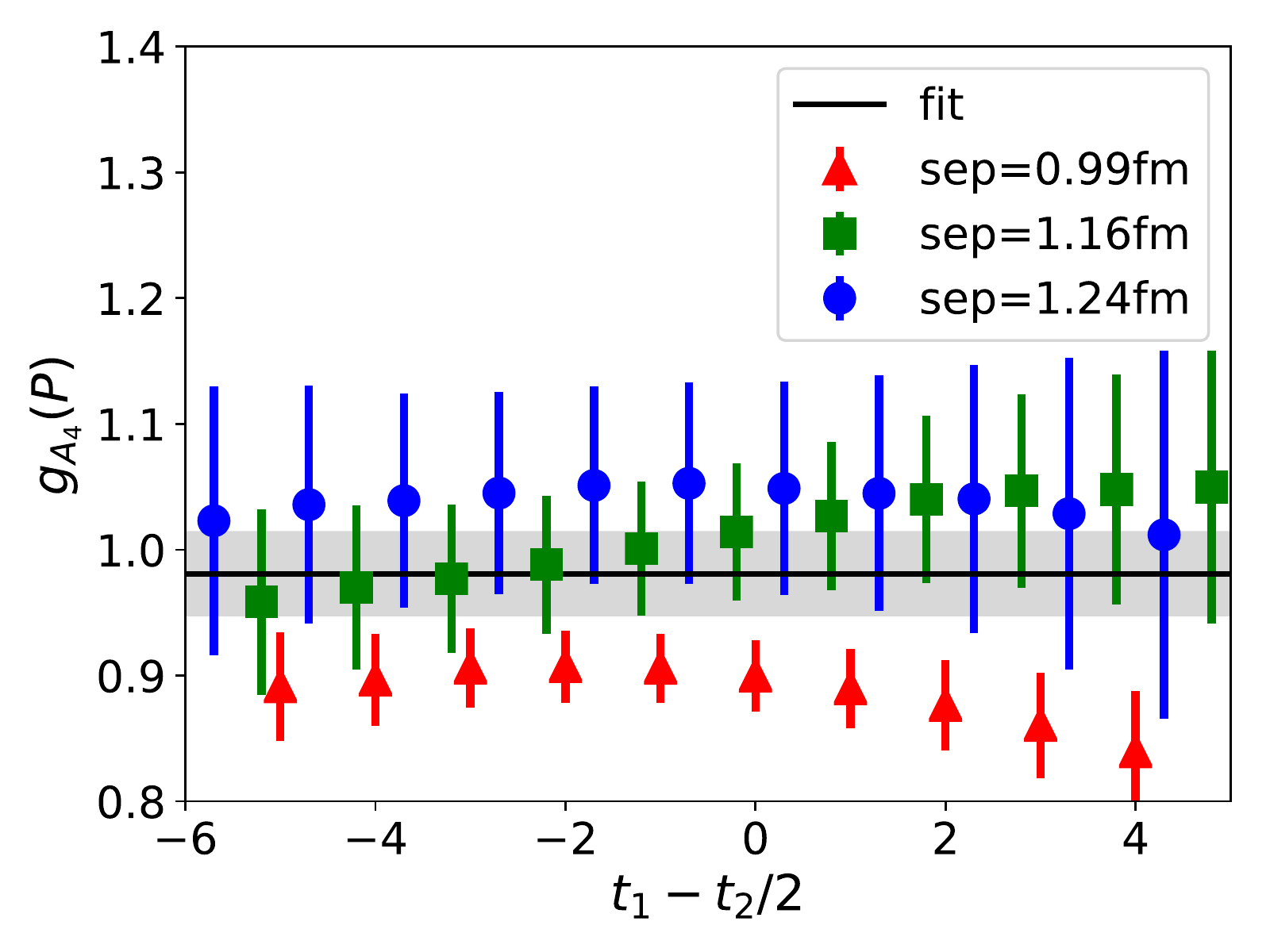}
\includegraphics[width=0.49\textwidth,page=1]{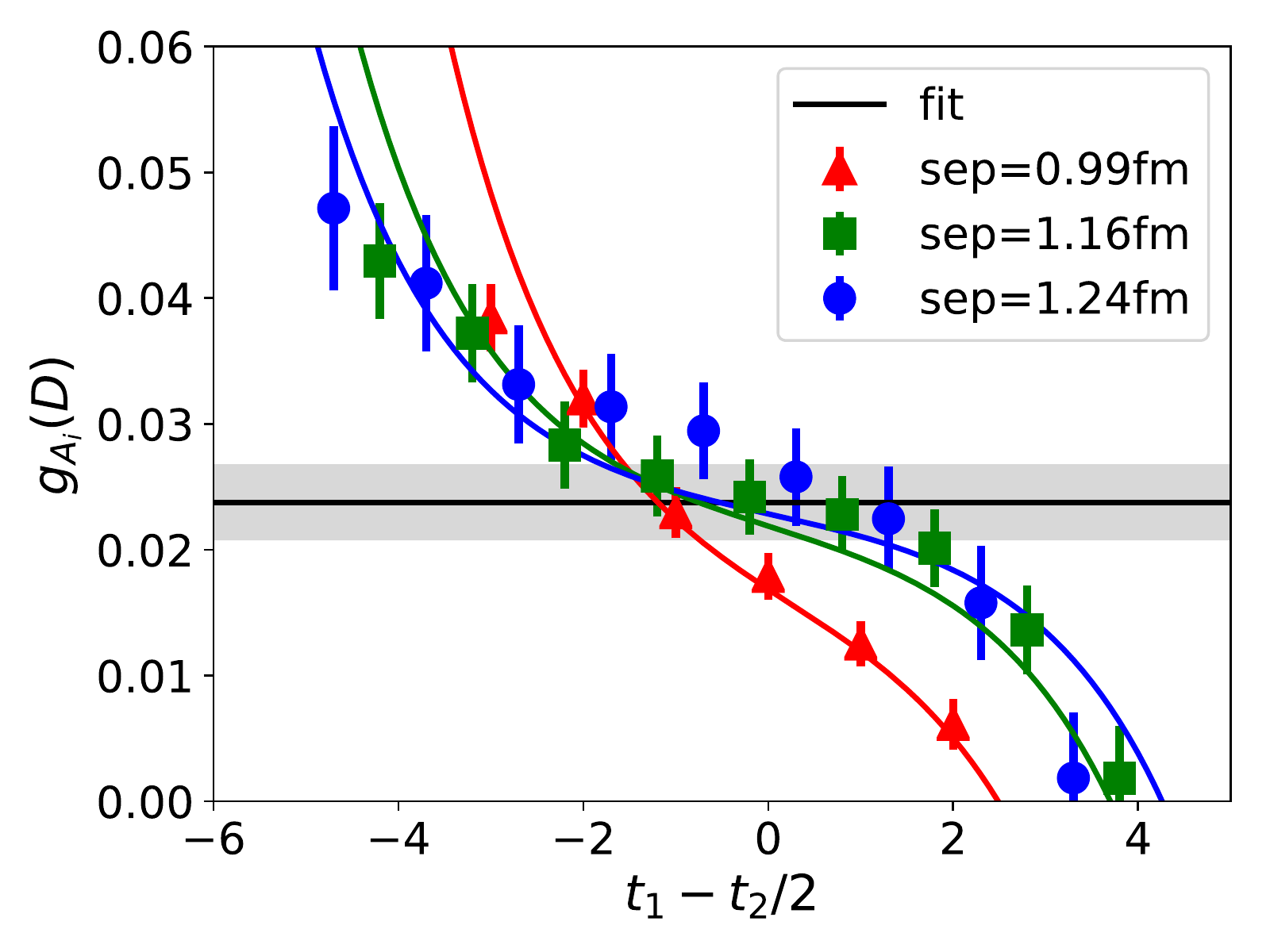}
\includegraphics[width=0.49\textwidth,page=1]{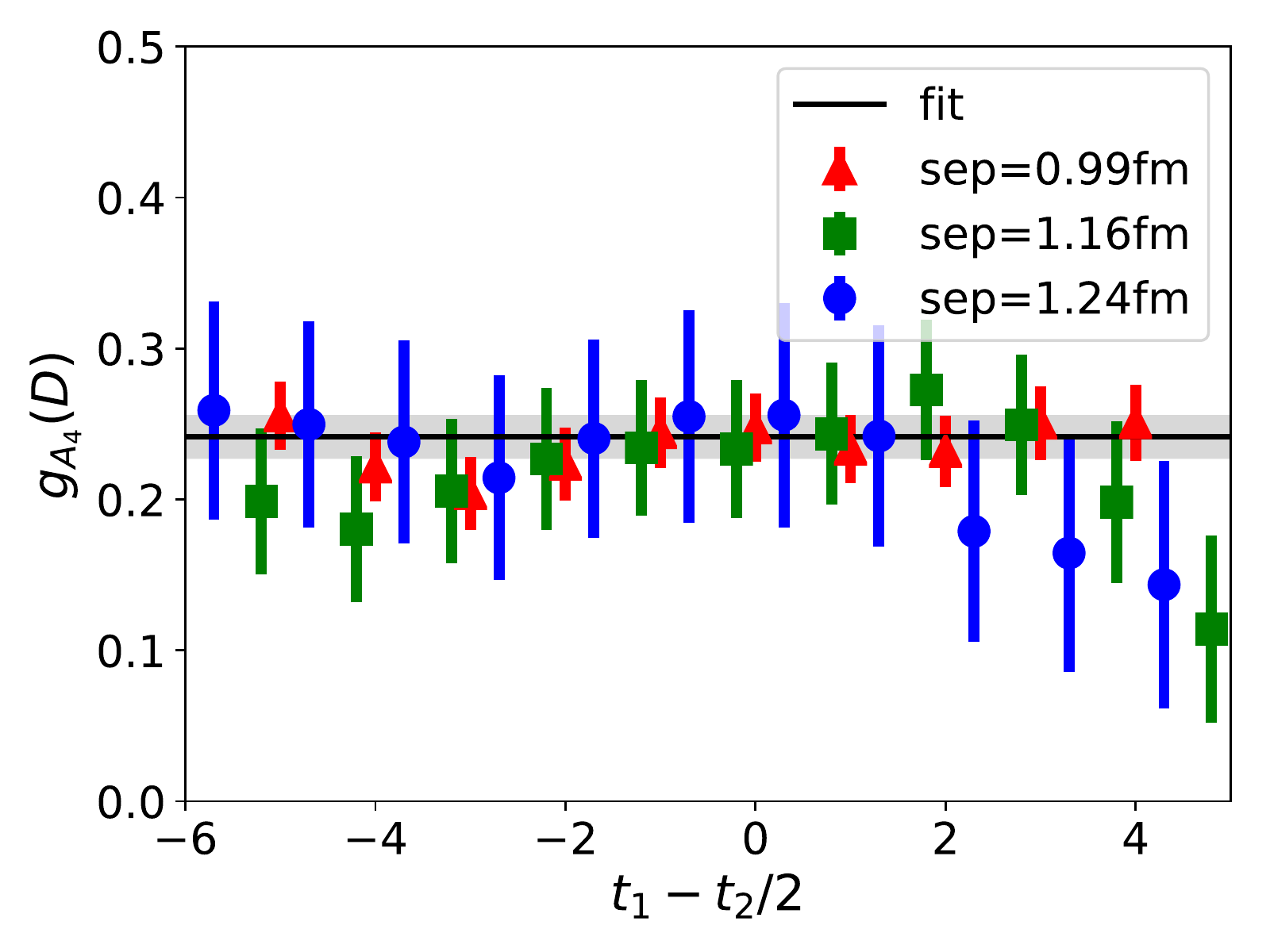}
\caption{2-state fits of 32I at the unitary point for the axial case.
(Points with error bars
are from lattice data and {the} curves are from the 2-state fits.) For different cases we keep different terms in the 2-state fit.
The gray bands indicate the final fit results.}
\label{32I_gA_m3}
\end{figure}
\begin{figure}[!h]
\includegraphics[width=0.49\textwidth,page=1]{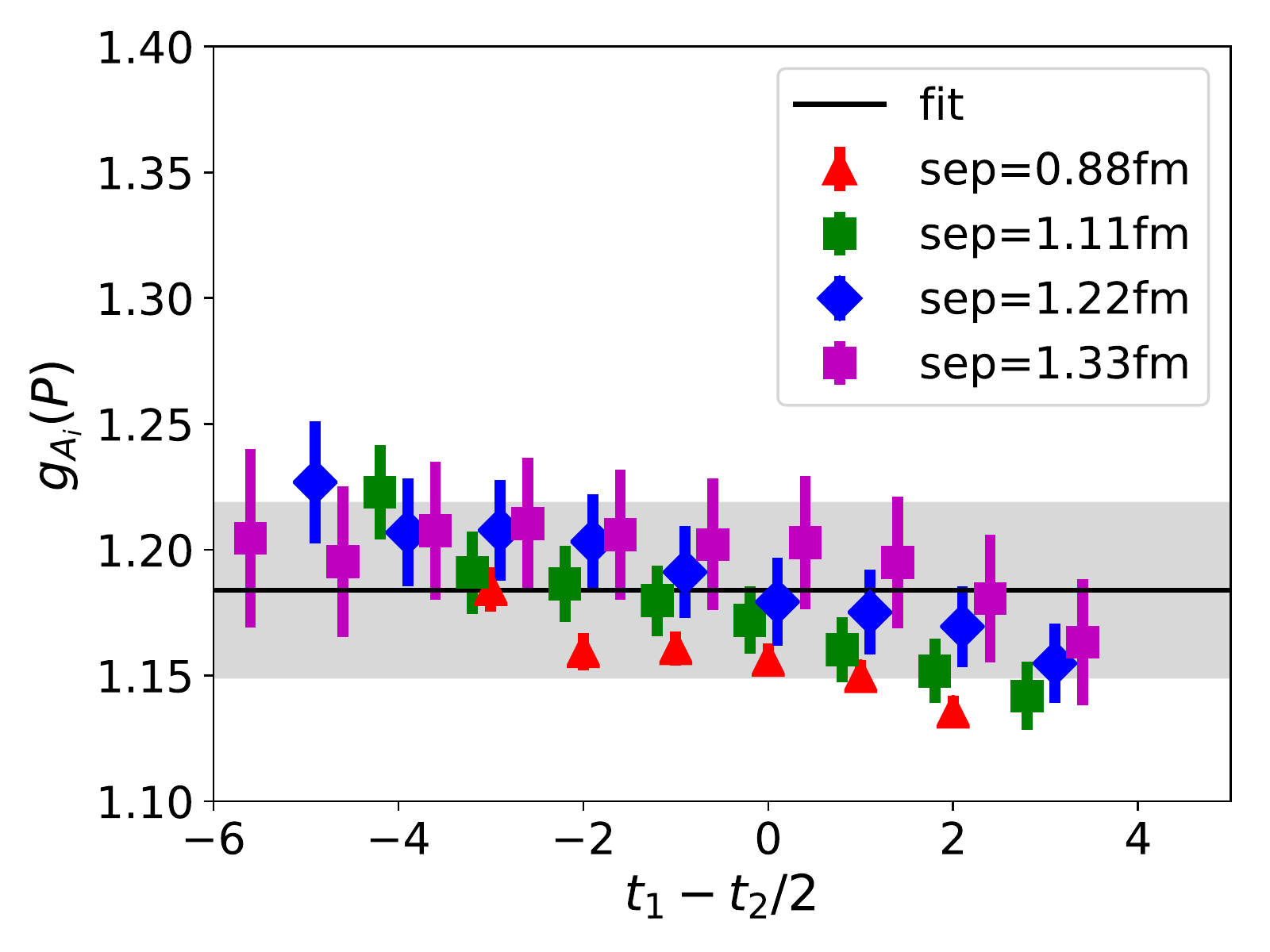}
\includegraphics[width=0.49\textwidth,page=1]{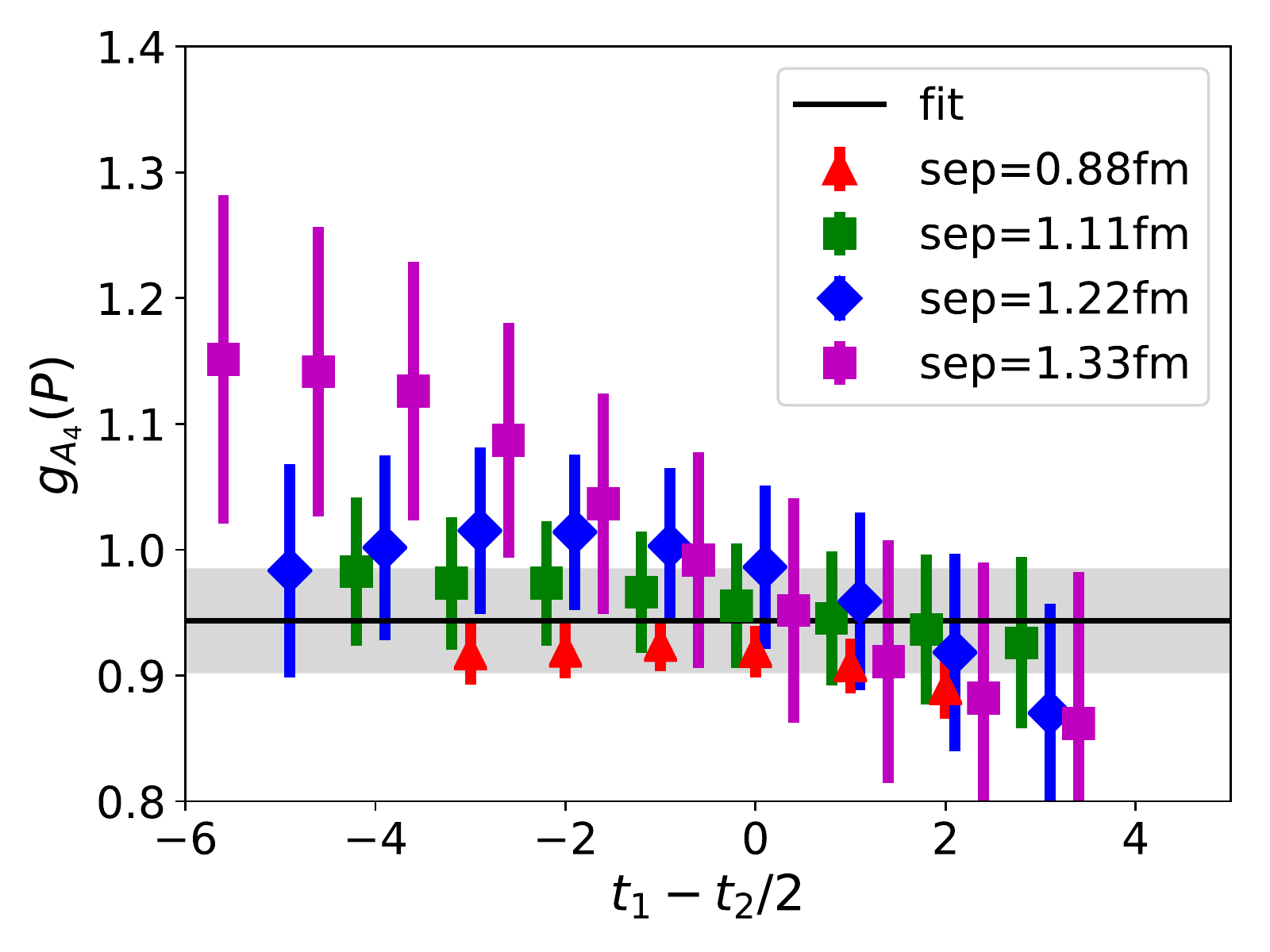}
\includegraphics[width=0.49\textwidth,page=1]{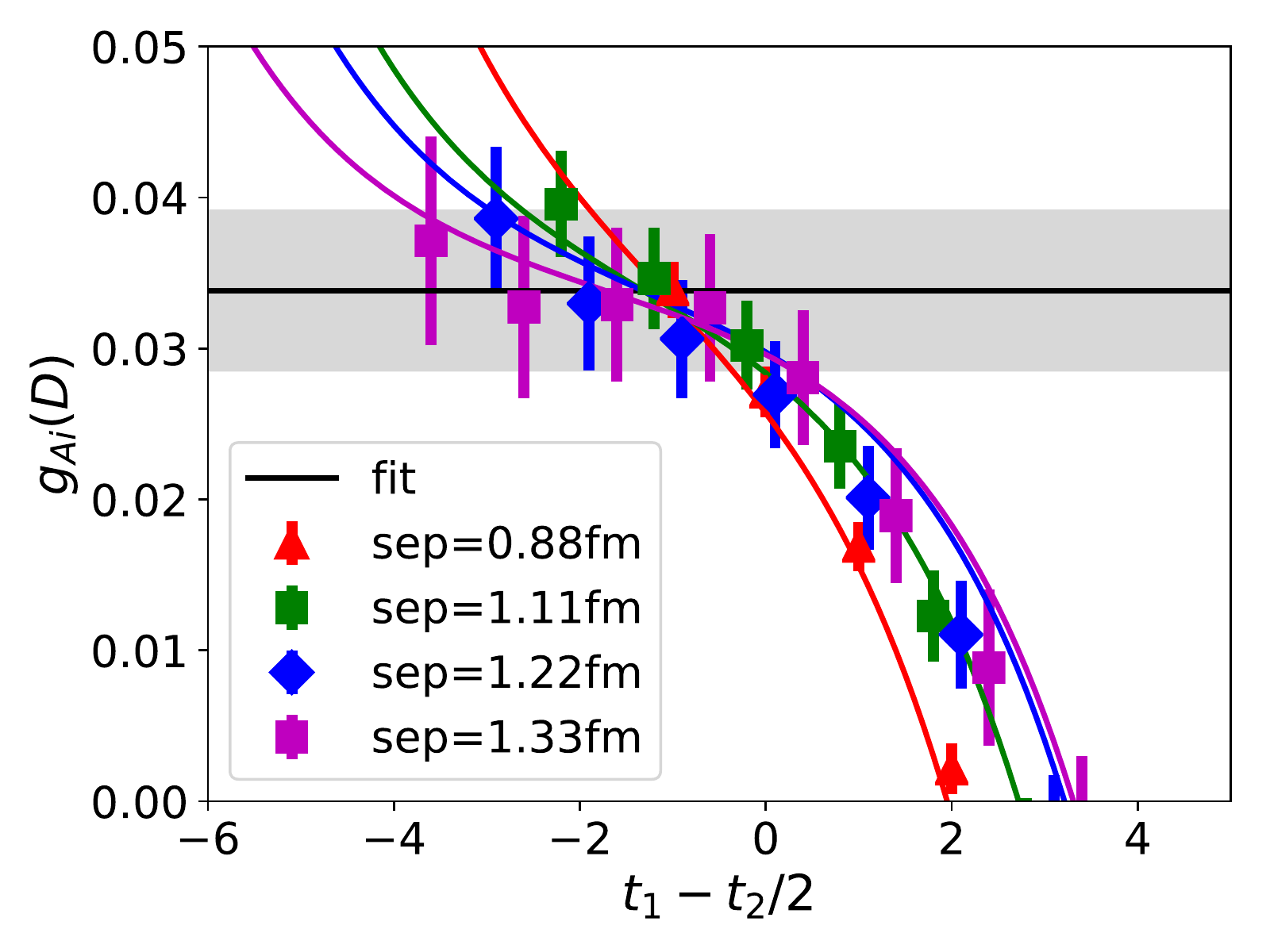}
\includegraphics[width=0.49\textwidth,page=1]{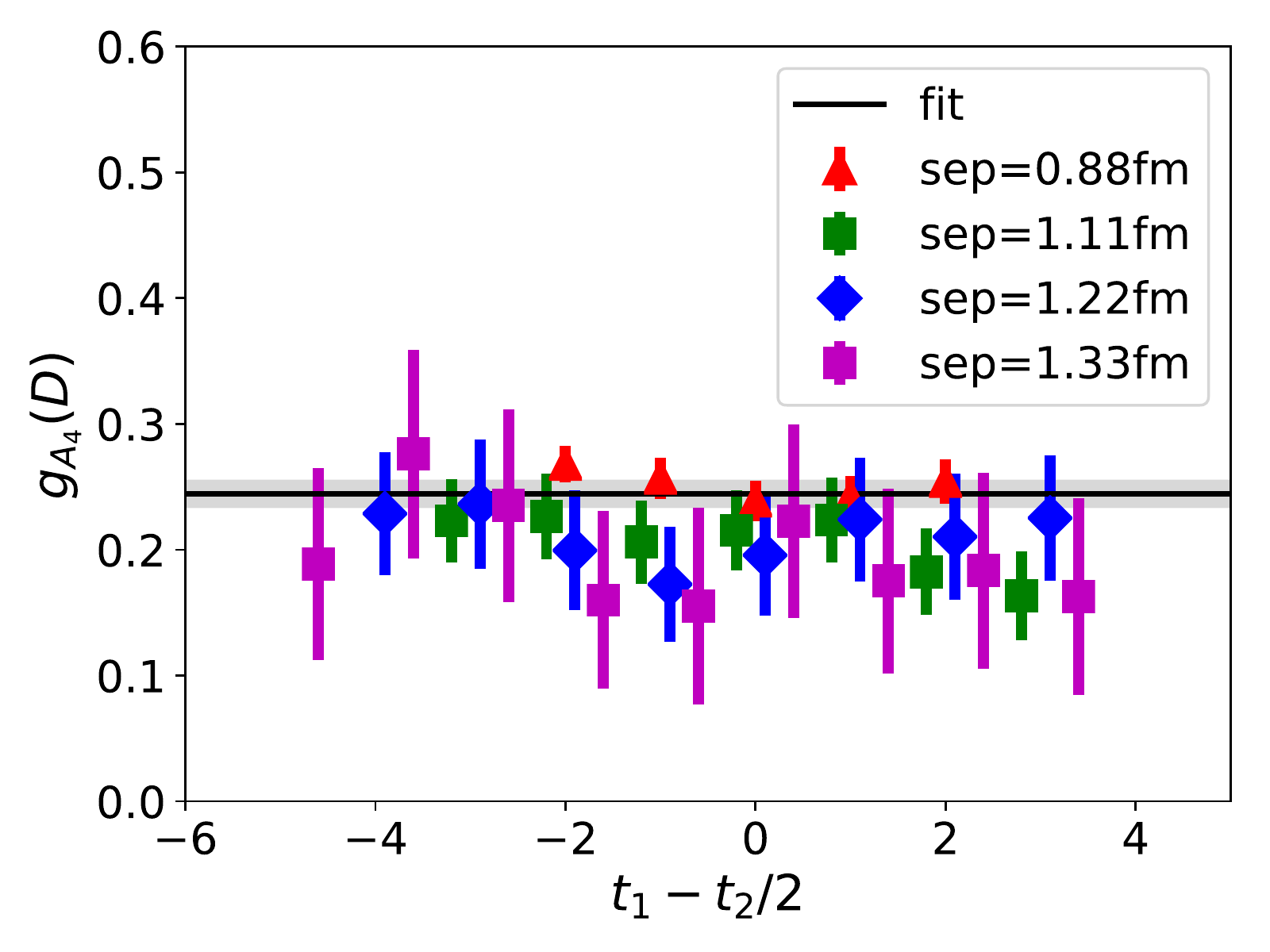}
\caption{The same as in Fig.~\ref{32I_gA_m3} but for lattice 24I at the unitary point.}
\label{24I_gA_m3}
\end{figure}

For the axial vector, we also try to use the standard 2-state fit to handle the excited-state contamination.
However, for $g_{A_4}(P)$, the lattice data points show no obvious curvature at either the source side or the
sink side on both lattices 24I and 32I within errors (the fit examples at the unitary points are shown in Fig.~\ref{32I_gA_m3} and \ref{24I_gA_m3}, respectively),
which means the transition terms are heavily suppressed.
If we still force a full 2-state fit on the data, the final uncertainty will be uncontrollable since the data have no constraint on the $C_1$ and $C_2$ terms.
But the data at different source-sink separations have some discrepancy, so we need to retain the $C_3$ term in addition to the constant $C_0$.
For $g_{A_i}(P)$, we utilize $C_0$ and both the $C_1$ and $C_3$ terms to fit the data.
In the practical fit procedure, we combine $g_{A_4}(P)$ and $g_{A_i}(P)$ into a joint fit with shared parameter $\delta m$,
where a wide prior $\delta m=0.3/a\pm0.3/a$ is used in some channels to ensure stable fit results.
For $g_{A_4}(D)$, we merely use a constant fit on the combination of three time separations.
For $g_{A_{i}}(D)$, although there is no evident plateau (for the same reason as for $g_{V_{4}}(D)$), a 2-state fit with the $C_3$ term fits the data well.
We can see that $g_{A_{4}}(D)$ is larger than $g_{A_{i}}(D)$ by a factor of $\sim7$.

\begin{figure}[!h]
\includegraphics[width=0.49\textwidth,page=1]{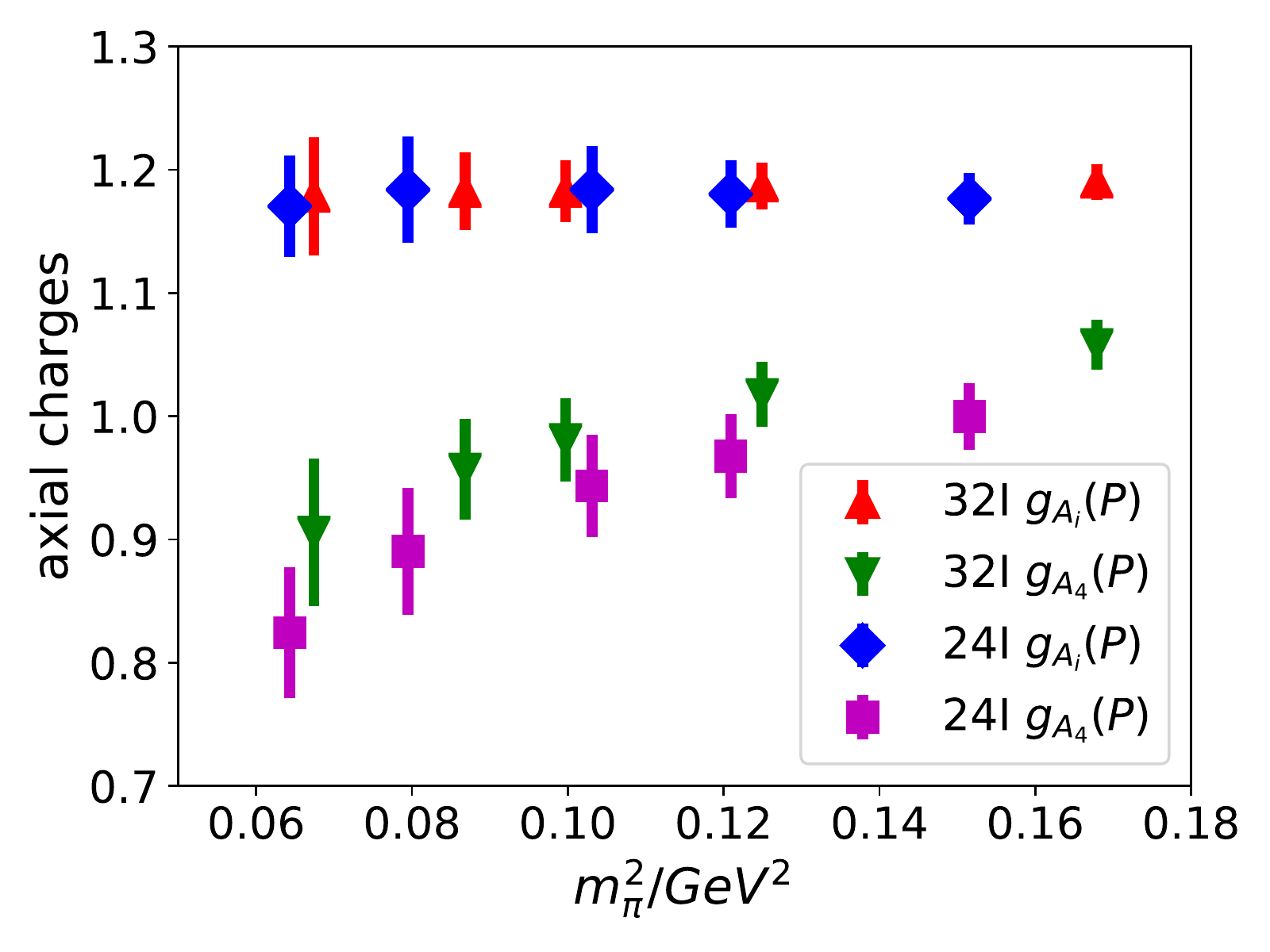}
\caption{Results of $g_{A_i}(P)$ and $g_{A_4}(P)$ as a function of squared pion mass, for both 24I and 32I.}
\label{32I24I_ga}
\end{figure}

The valence pion mass dependence of the axial vector charges can be found in Fig.~\ref{32I24I_ga}. The values are all
normalized.
$g_{A_i}(P)$ has hardly any pion mass dependence, which is consistent with other calculations~\cite{Alexandrou:2013joa,Bhattacharya:2013ehc},
and the results from 24I and 32I are consistent with each other.
At the unitary point, we have $g_{A_i}(P)(24I)=1.18(4)$ and $g_{A_i}(P)(32I)=1.19(3)$, which are lower than the
experimental value, and the deviation is around 7 percent.
This is also consistent with other calculations~\cite{Alexandrou:2013joa,Owen:2012ts,Bhattacharya:2016zcn} for pion mass $\sim300$ MeV.
On the other hand, $g_{A_4}(P)$ is smaller than $g_{A_i}(P)$ by $\sim20\%$,
and it decreases with decreasing pion mass.
At the unitary point, the gap between $g_{A_i}(P)$ and $g_{A_4}(P)$ is around $7 \sigma$.
This deviation gets smaller with increasing pion mass.
If this behavior continues, the gap will vanish in the strange
quark region. 
Actually, we have done some rough calculations with strange quark mass
and we do not see any discrepancy there within errors.
This is evidence that the deviation has a physical origin and is
not a mere statistical fluctuation.

\begin{figure}[!h]
\includegraphics[width=0.49\textwidth,page=1]{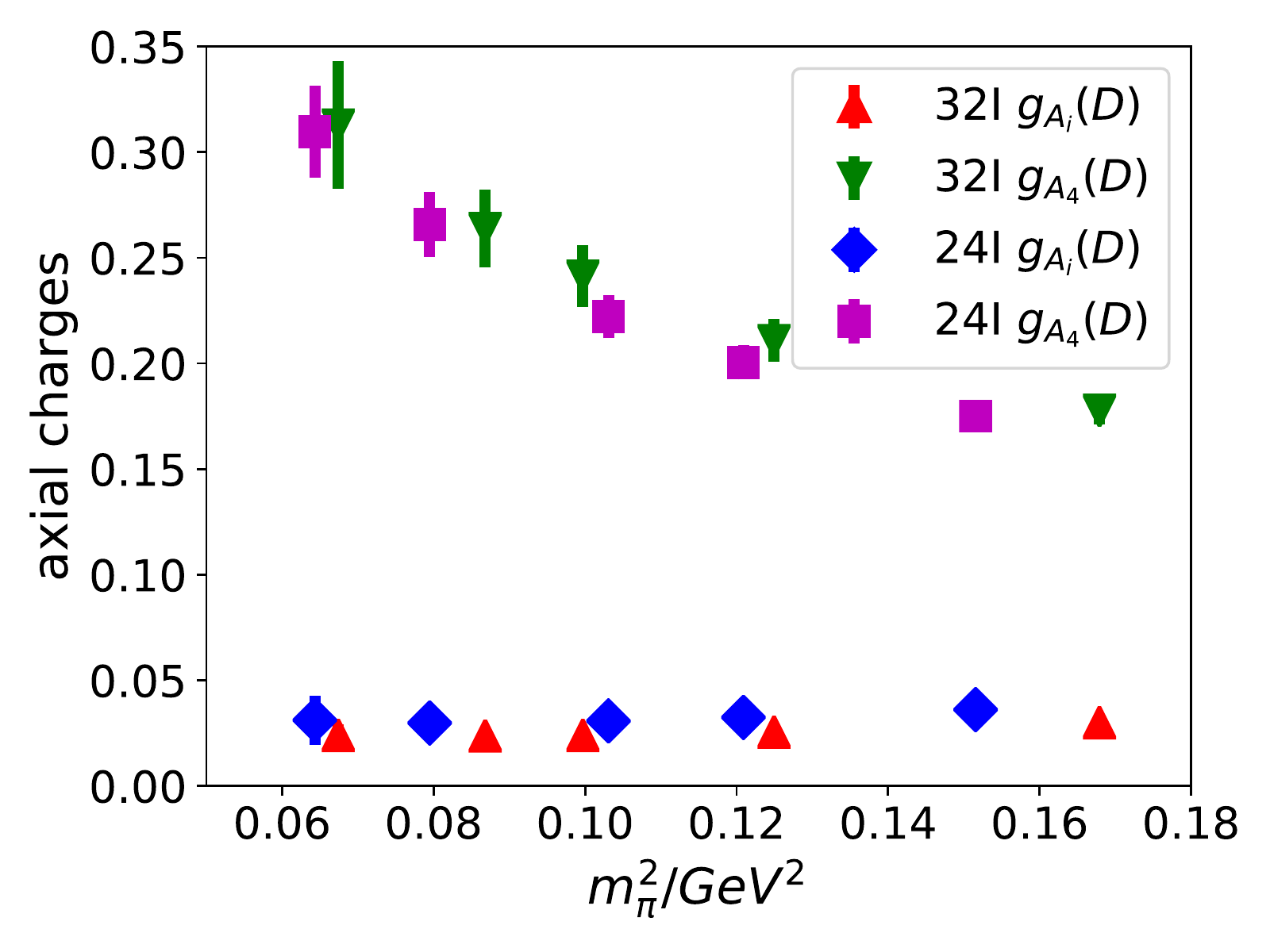}
\caption{Results of the improvement currents
$g_{A_{i}}(D)$ and $g_{A_{4}}(D)$ as a function of pion mass squared, for 24I and 32I.}
\label{32I24I_Ad}
\end{figure}

\begin{figure}[!h]
\includegraphics[width=0.49\textwidth,page=1]{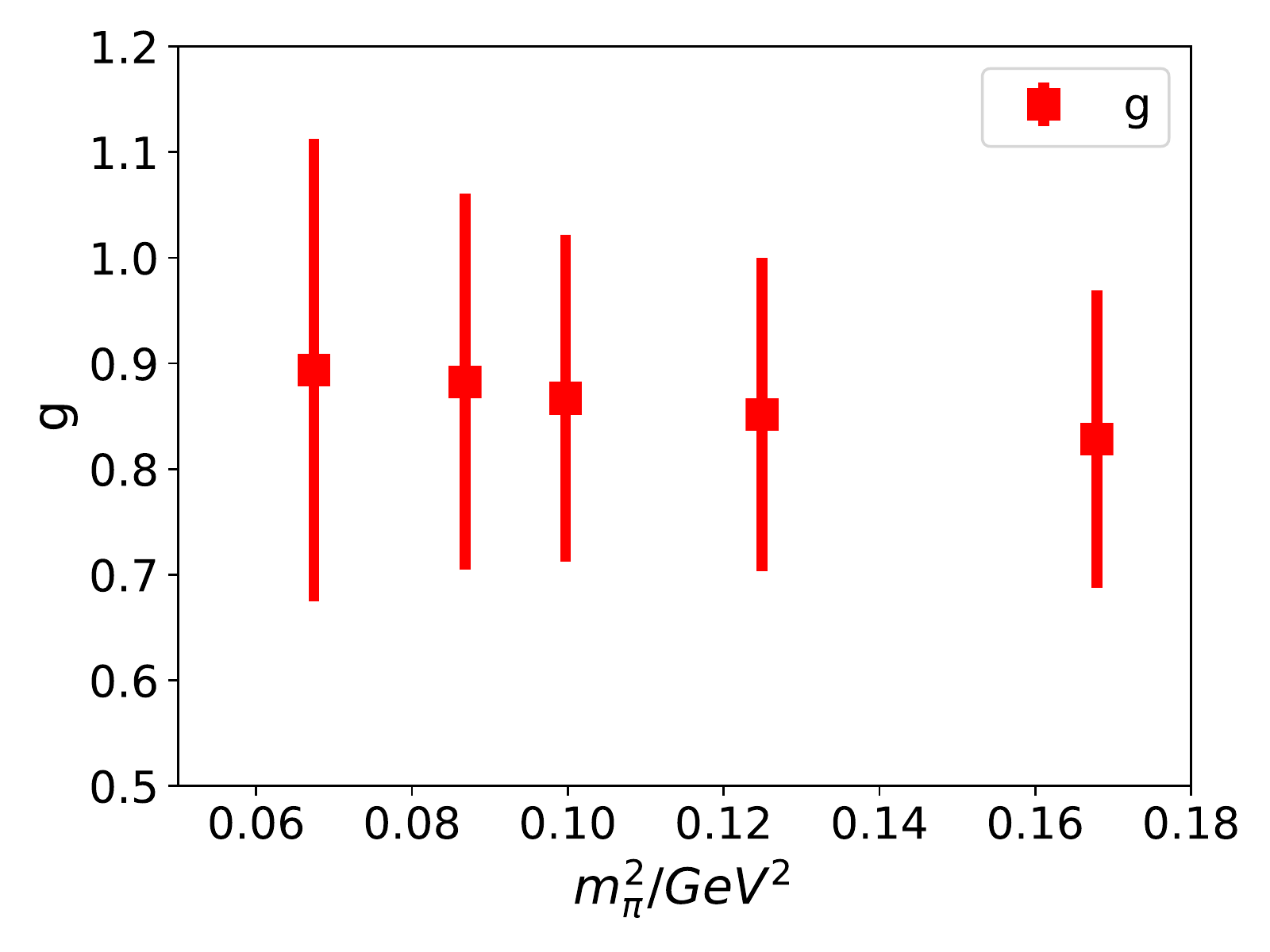}
\caption{The coefficient $g$ solved on 32I. The errors of $g$ come from bootstrap.}
\label{32I_f}
\end{figure}

\begin{figure}[!h]
\includegraphics[width=0.49\textwidth,page=1]{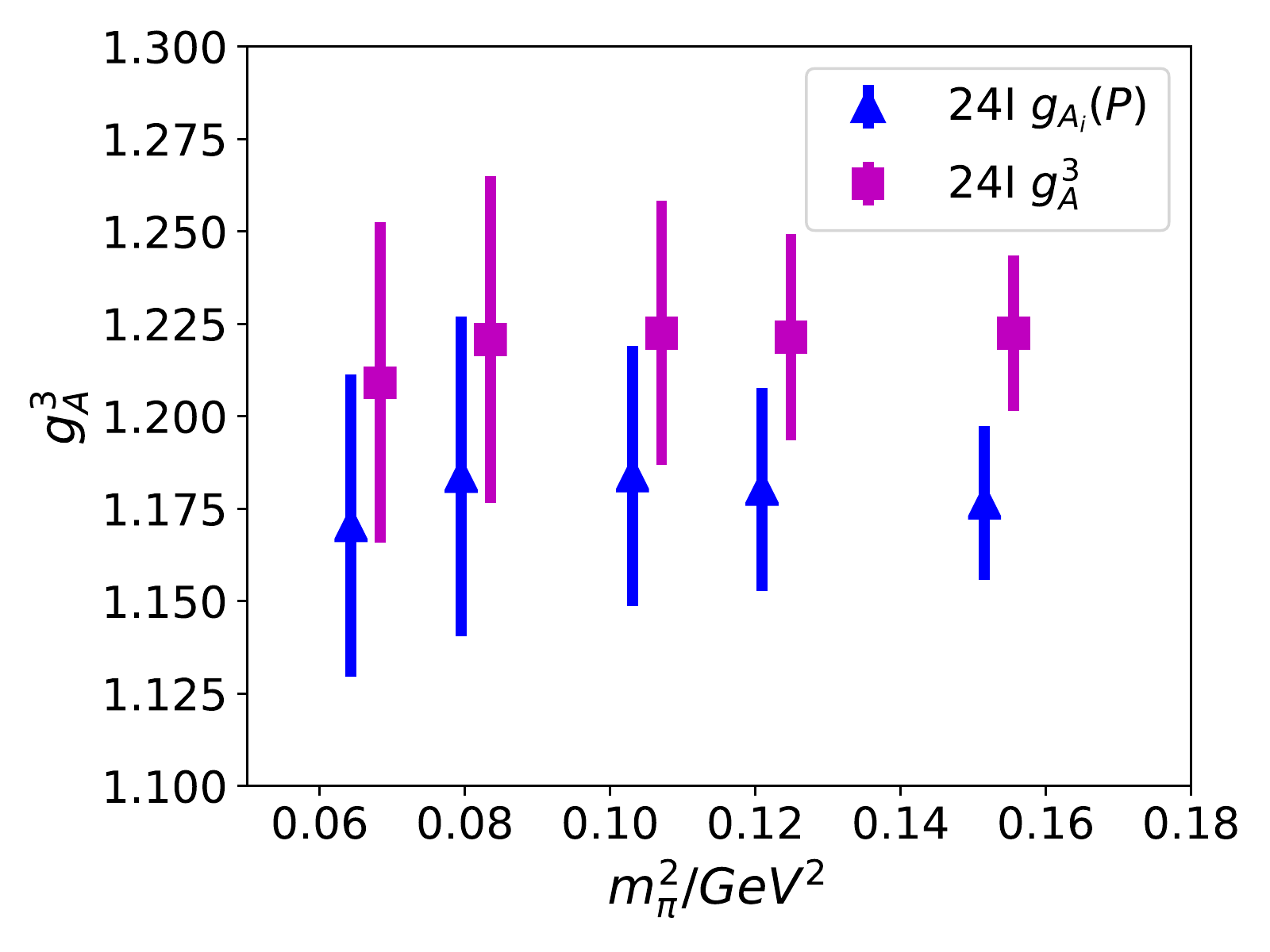}
\includegraphics[width=0.49\textwidth,page=1]{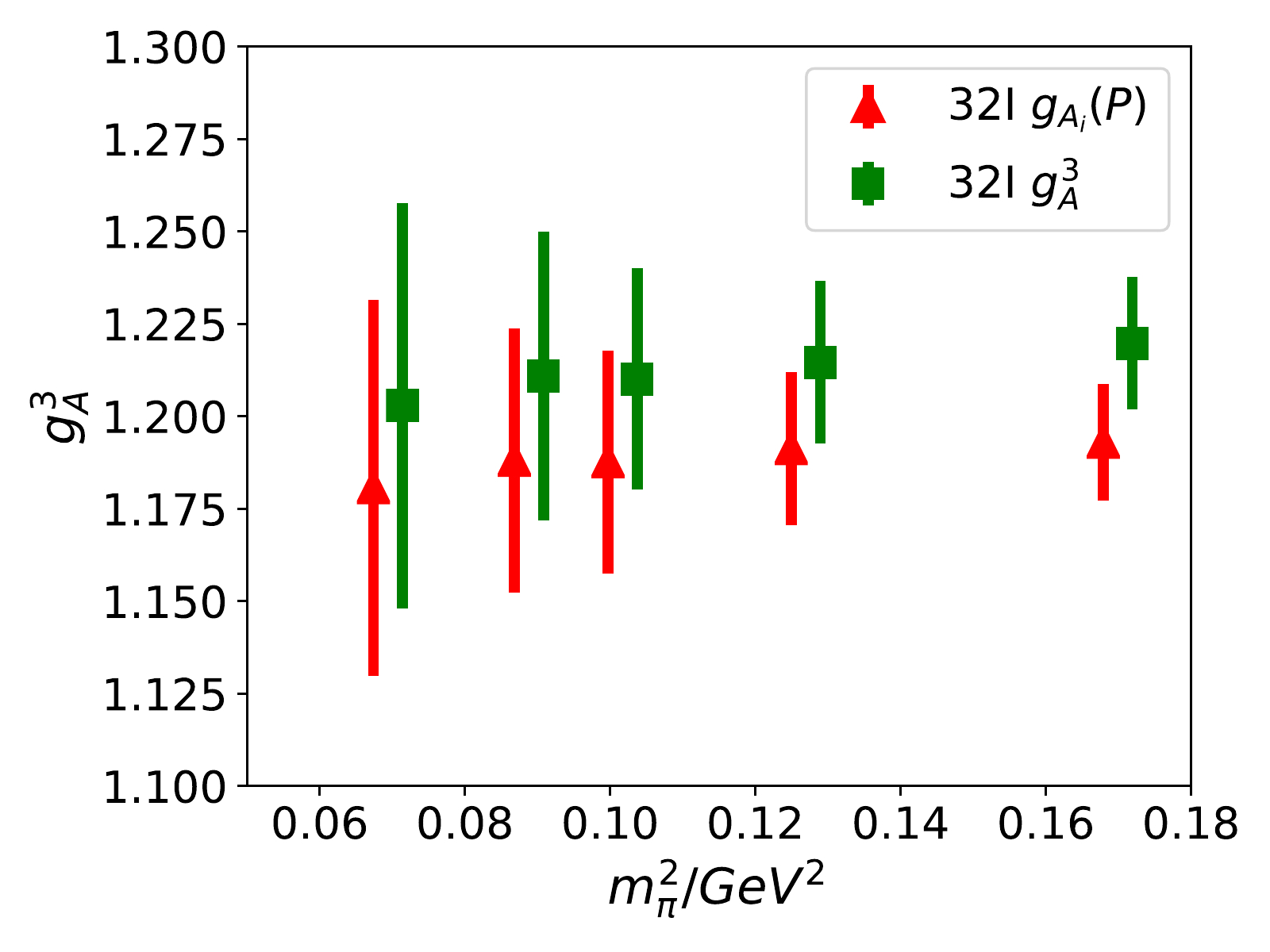}
\caption{The final improved axial charge values of 24I and 32I compared with the unimproved $g_{A_i}(P)$.}
\label{32I24I_im}
\end{figure}

It is very interesting that the currents with derivatives (see Fig.~\ref{32I24I_Ad}) manifest exactly the opposite
behavior to those of the point currents. This time $g_{A_{4}}(D)$ has strong $m_{\pi}$ dependence and the value increases as the pion mass decreases, 
while for $g_{A_{i}}(D)$,
the central values are tiny and almost constant within errors.
These opposite behaviors between the dimension-3 local currents and the dimension-4 derivative currents is exactly the pattern that is needed to
implement the improvement and which makes the improved $g_A(P)$ larger and thus closer to the experimental value.
Using Eq.~(\ref{g and f}),
we can indeed obtain the coefficient $g$ at each pion mass and, furthermore, the value of $g$
is constant ($\sim0.85$ for 32I and $\sim1.14$ for 24I) at different pion masses within errors, 
which is what we expect (since mass dependence entails a higher dimensional correction) and is shown in Fig.~\ref{32I_f}.
The errors of $g$ come from bootstrap resampling: we do 2-state fit (or constant fit) and solve Eq.~(\ref{g and f})  
in each bootstrap sample.
Since the coefficient $g$ is positive and the charges related to the currents with derivative operators are also positive,
the final improved axial charges will be larger than the original values.
The improved results are presented in Fig.~\ref{32I24I_im}.
At the unitary point, the improved axial charges are $g_A^{3}(24I)=1.22(4)$ and $g_A^{3}(32I)=1.21(3)$,
which represents a 3.4\% and 1.7\% increment towards the experimental value, respectively.
For other pion masses, there are $2\%\sim4\%$ improvements as well.

The fit ranges of each matrix elements are listed in Table~(\ref{fit_windows}) and
all the fitting results at different values of pion mass are presented in Table~(\ref{fit_results}) for further reference.
Since this is an exploratory work for addressing the possible solution of the discrepancy of $g_{A_4}(p)$ and $g_{A_i}(p)$
and the sea pion mass used is far away from the physical point,
we do not carry out a chiral and continuum extrapolation to get the physical value of the axial charge.
The systematic uncertainties of the $g_A^{3}$ obtained at the unitary points mainly come from the fit of the
3-point function to 2-point function ratios and the improvement scheme we are using.
The fit of $g_{A_i}(p)$ is stable since the data points are precise and we only used $C_0$ and the $C_3$ terms to do the fitting.
Choosing different fitting windows can result in $\sim2\%$ difference which can be treated as a systematic uncertainty
in the fit of $g_{A_i}(p)$. The fits of other matrix elements are not as stable as the $g_{A_i}(p)$ case, but these matrix elements are only
used to calculate the improvement. Since the improvement itself is around $3\%$, even if the matrix element shifts by $50\%$, the final uncertainty of the
improved value caused by this will be only $1.5\%$. This is the second part of the systematic uncertainties.
The systematic uncertainty of our improvement scheme is hard to estimate, but for the same reason that the improvement is only around $3\%$,
the uncertainty of the final value induced by our scheme should not be larger that  $1\%$.
So the total systematic uncertainty will be $\sim2.7\%$.
The improved axial charges at the unitary point with systematic errors are {$g_A^{3}(24I)=1.22(4)(3)$ and $g_A^{3}(32I)=1.21(3)(3)$}.

\begin{table}[!h]
\begin{center}
\begin{tabular}{cccccccccccccccccc}
Lattice &  $g_{A_i}(P)$  & $g_{A_4}(P)$ & $g_{V_i}(P)$ & $g_{V_4}(P)$ & $g_{A_i}(D)$  & $g_{A_4}(D)$ & $g_{V_i}(D)$ & $g_{V_4}(D)$\\
\hline
24I  &  4   &  4  &  2  &  4  &  3   &  2  &  4  &  4 \\
32I  &  5   &  5  &  5  &  5  &  4   &  5  &  3  &  3 \\
\hline
\end{tabular}
\end{center}
\caption{The number of data points dropped on both the source and the sink side of 3-point function to 2-point function ratios when doing the fit for the
corresponding matrix elements.}
\label{fit_windows}
\end{table}

\begin{table}[!h]
\begin{center}
\begin{tabular}{cccccccccccccccccc}
Lattice & $m_{\pi}$(MeV) & $m_N$(GeV) & $g_{A_i}(P)/\delta_ma$  & $g_{A_4}(P)/\delta_ma$ & $g_{V_i}(P)/\delta_ma$ & $g_{V_4}(P)$ & g\\ \cline{4-7}
 & & & $g_{A_i}(D)/\delta_ma$  & $g_{A_4}(D)$ & $g_{V_i}(D)/\delta_ma$ & $g_{V_4}(D)/\delta_ma$ & \\ 
\hline
\hline
24I & 254 & 1.083(8) & 1.18(4)/0.27(30) & 0.83(5)/0.27(30) & 1.19(28)/0.30(23) & 1.001(5) & 1.13(32) \\ \cline{4-7}
      &        &                & 0.03(1)/0.5(2) & 0.31(2) &  0.02(2)/0.5(4.3)                            &  0.007(12)/2.2(1.5)&\\
\hline
24I & 282 & 1.102(6) & 1.19(4)/0.23(30) &  0.90(5)/0.23(30) & 1.18(23)/0.28(18) & 1.000(4)  & 1.13(27) \\ \cline{4-7}
      &        &                & 0.030(8)/0.5(2)& 0.27(1) &   0.02(1)/0.5(3.8)                            &  0.012(8)/1.7(6)               &\\
\hline
24I & 321 & 1.131(5) & 1.19(4)/0.25(30) & 0.95(4)/0.25(30) & 1.16(18)/0.26(14) & 0.999(3)  & 1.14(21) \\ \cline{4-7}
      &        &                & 0.031(5)/0.6(2)& 0.23(1) &   0.025(10)/0.6(3.4)                           &  0.017(6)/1.2(3)               &\\
\hline
24I & 348 & 1.152(4) & 1.19(3)/0.27(30) & 0.98(3)/0.27(30) & 1.14(15)/0.27(13) & 1.000(3)  & 1.15(20) \\ \cline{4-7}
      &        &                & 0.033(4)0.6(1)& 0.21(1) &   0.027(7)/0.7(3.0)                           &  0.021(5)/0.97(18)               &\\
\hline
24I & 389 & 1.196(4) & 1.19(2)/0.29(29) & 1.11(3)/0.29(29) & 1.11(11)/0.29(13) & 0.999(3)  & 1.16(18) \\ \cline{4-7}
      &        &                & 0.037(3)/0.6(1)& 0.184(5)&   0.030(5)/0.8(2.5)                         & 0.026(4)/0.74(13)                &\\
\hline
32I & 260 & 1.087(5) & 1.18(5)/0.3(3) & 0.91(6)/0.3(3) & 1.5(7)/0.14(8) & 1.012(6) & 0.89(22) \\ \cline{4-7}
      &        &                & 0.024(5)/0.7(2) & 0.35(4) &  0.02(2)/0.4(1.2)                            &  0.02(1)/0.38(7)&\\
\hline
32I & 295 & 1.112(4) & 1.18(3)/0.3(3) & 0.96(4)/0.3(3) & 1.3(3)/0.17(9) & 1.011(4) & 0.88(18) \\ \cline{4-7}
      &        &                & 0.024(4)/0.7(1) & 0.28(2) &  0.02(1)/0.4(0.6)                            &  0.024(8)/0.36(4)&\\
\hline
32I & 316 & 1.128(4) & 1.18(2)/0.4(3) & 0.98(3)/0.4(3) & 1.1(2)/0.13(8) & 1.011(4) & 0.87(15) \\ \cline{4-7}
      &        &                & 0.024(3)/0.7(1) & 0.25(2) & 0.024(9)/0.5(0.5)                           &  0.028(6)/0.35(4)&\\
\hline
32I & 353 & 1.156(3) & 1.19(2)/0.4(2) & 1.02(3)/0.4(2) & 1.04(5)/0.15(8) & 1.011(3) & 0.85(15) \\ \cline{4-7}
      &        &                & 0.025(2)/0.67(8) & 0.21(1) &  0.026(5)/0.6(0.4)                         &  0.033(4)/0.34(3)&\\
\hline
32I & 410 & 1.208(2) & 1.19(1)/0.4(2) & 1.06(2)/0.4(2) & 1.01(2)/0.41(8) & 1.010(2) & 0.83(14) \\ \cline{4-7}
      &        &                & 0.030(2)/0.63(6) & 0.18(1) &  0.030(3)/0.7(0.4)                         &  0.038(3)/0.33(2)&\\
\hline
\end{tabular}
\end{center}
\caption{The fitting results of all the matrix elements, the nucleon mass and the factor $g$. 
The parameter $\delta_m$ which is the mass difference between the first excited state and the nucleon
is also listed for the channels using 2-state fit.}
\label{fit_results}
\end{table}

Although the $\sim3\%$ improvement is still not enough to fill the gap of $\sim7\%$ between lattice and experiments,
it at least is in the correct direction. The results here are all from lattices with $m_\pi\sim 300$MeV; for the results around the physical pion mass, 
this several percent of improvement may be more significant.
We believe that the long-standing deviation of $g^3_A$ from the experimental value 
is probably not due to one single source. It may be a combined effect
of finite lattice volume, finite lattice spacing, and excited-state contamination.
Although the improvement currents contributes only two or three percent, they should be taken into account at finite cut off before
approaching the continuum, in addition to the finite normalization factors $Z_V$ and $Z_A$.

\section{a test of clover case}
\label{a test of clover case}
As a benchmark test of our results and the implementation of our low-mode substitution sandwich method, 
we carry out a test calculation using clover fermions as valence on the 
same DWF sea on lattice 24I. We also want to know whether the discrepancy observed above is only an artifact of overlap fermion or a more common phenomenon.
We use the standard sink-sequential method for constructing the 3-point functions without any low-mode substitution. Therefore, the forward-backward derivative operator
$\overleftrightarrow{D}$ can be  implemented easily; we can therefore check the difference between using $\overleftrightarrow{D}$ and $\overrightarrow{D}$.
Moreover, we also try to figure out whether our improvement is still valid for each $u$ or $d$ flavor separately (connected insertion part only).
In this test,
the $C_{sw}$ is chosen to be the tadpole improved value which is $\frac{1}{u_0^3}$,
where the tadpole parameter $u_0$ is the fourth root of the plaquette. Since our configurations
are once HYP-smeared, the plaquette values are around 0.94, so $C_{sw}=\frac{1}{0.94^{0.75}}\sim1.05$.
The mass parameter $m=-0.058$ is chosen to make
the pion mass {to} be around $300$ MeV which is similar to the unitary point of the 24I lattice. 
We use only one source for the calculation and the source sink separation is fixed to be $8a$.

\begin{figure}[!h]
\includegraphics[width=0.49\textwidth,page=1]{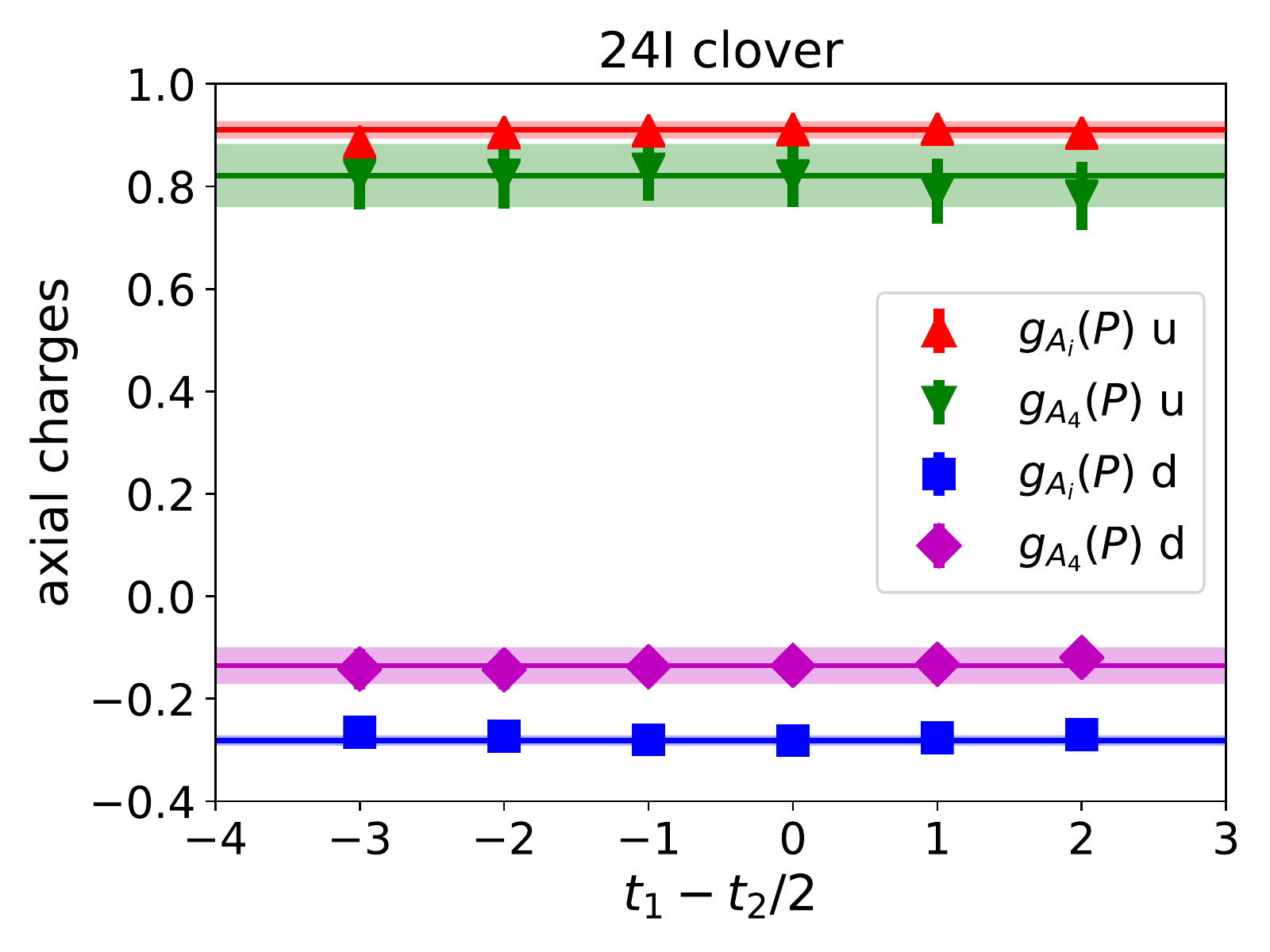}
\includegraphics[width=0.49\textwidth,page=1]{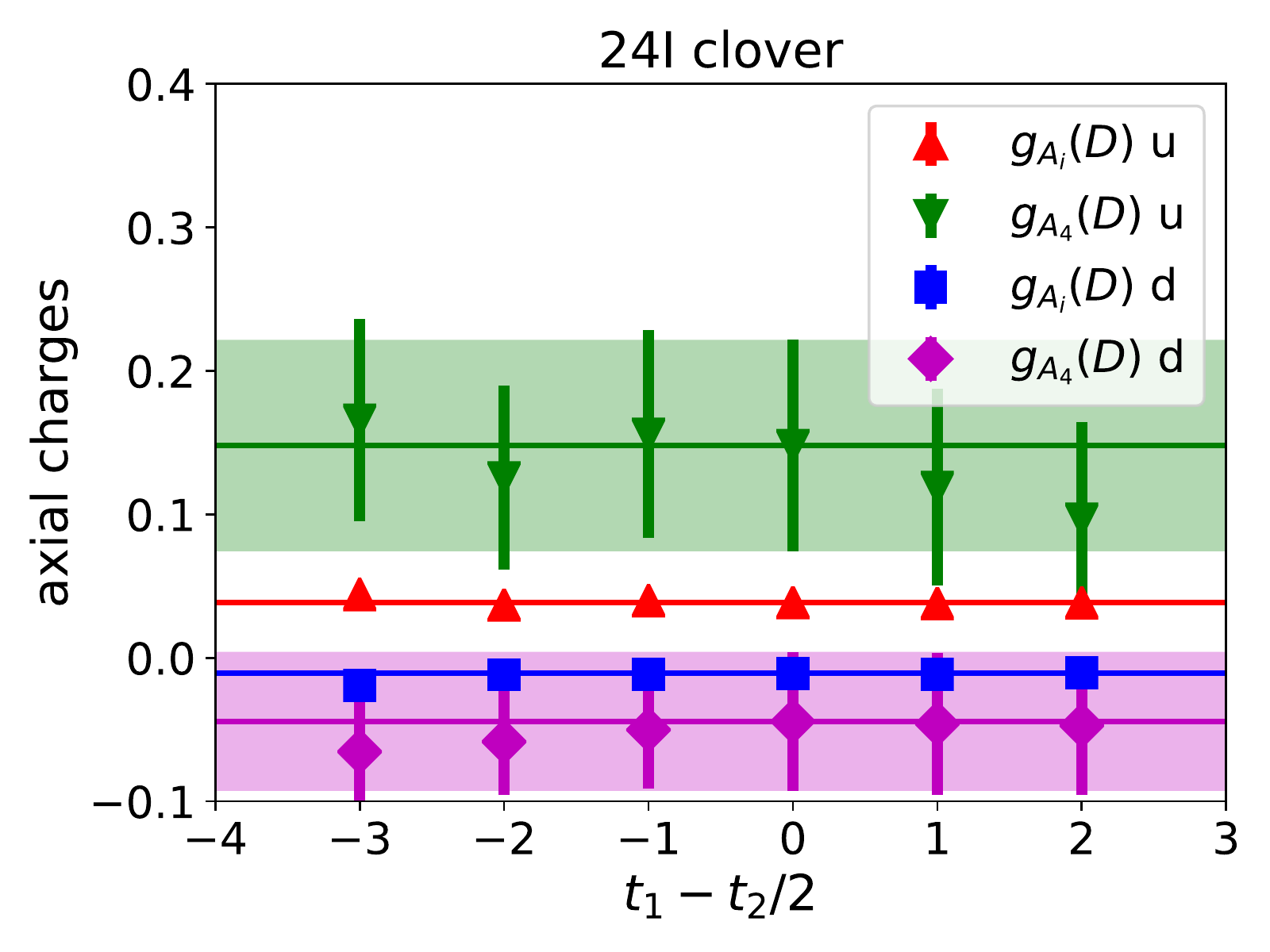}
\caption{The results of the clover test for separate $u$ and $d$.}
\label{clover_test}
\end{figure}

\begin{figure}[!h]
\includegraphics[width=0.49\textwidth,page=1]{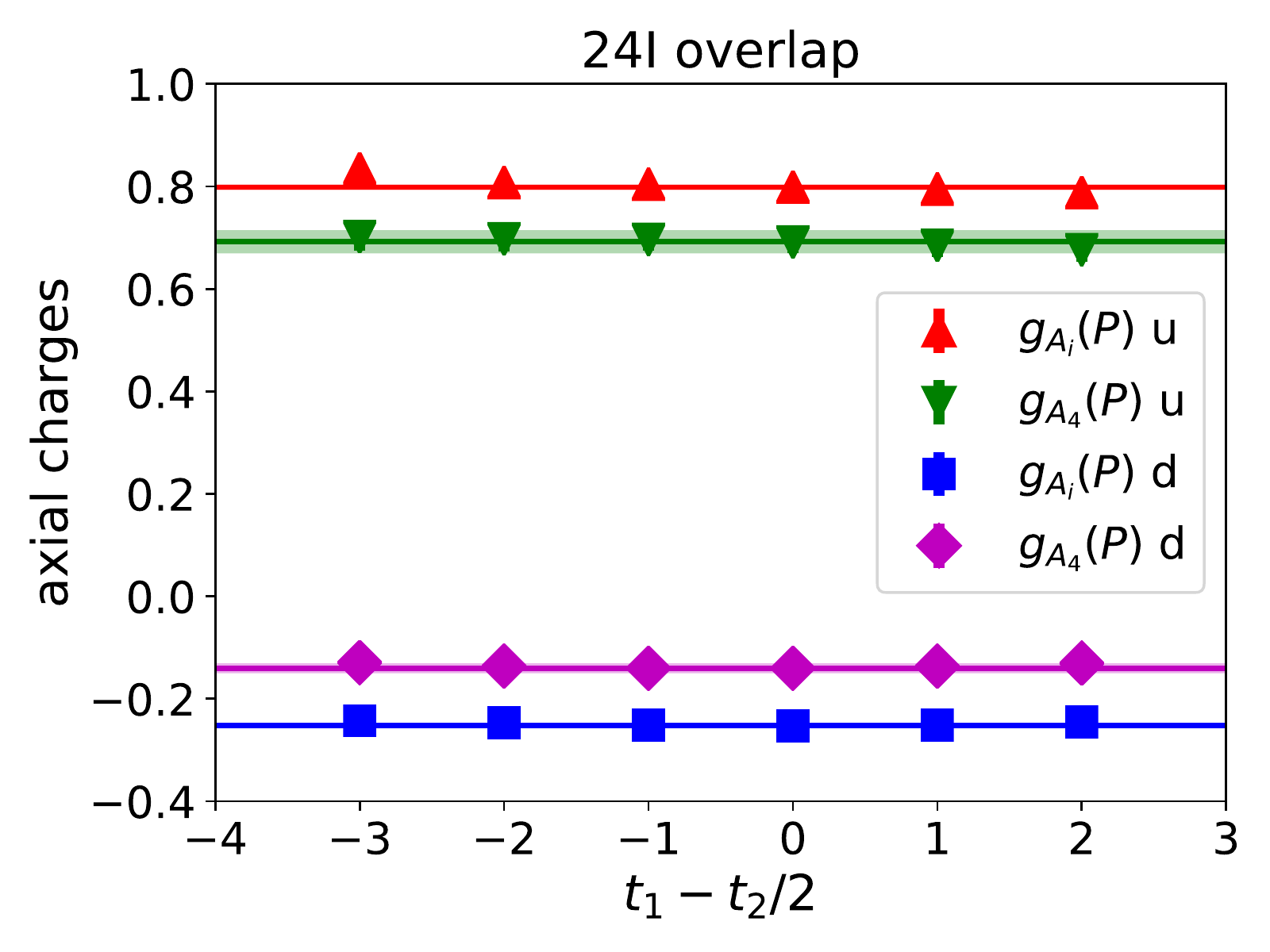}
\includegraphics[width=0.49\textwidth,page=1]{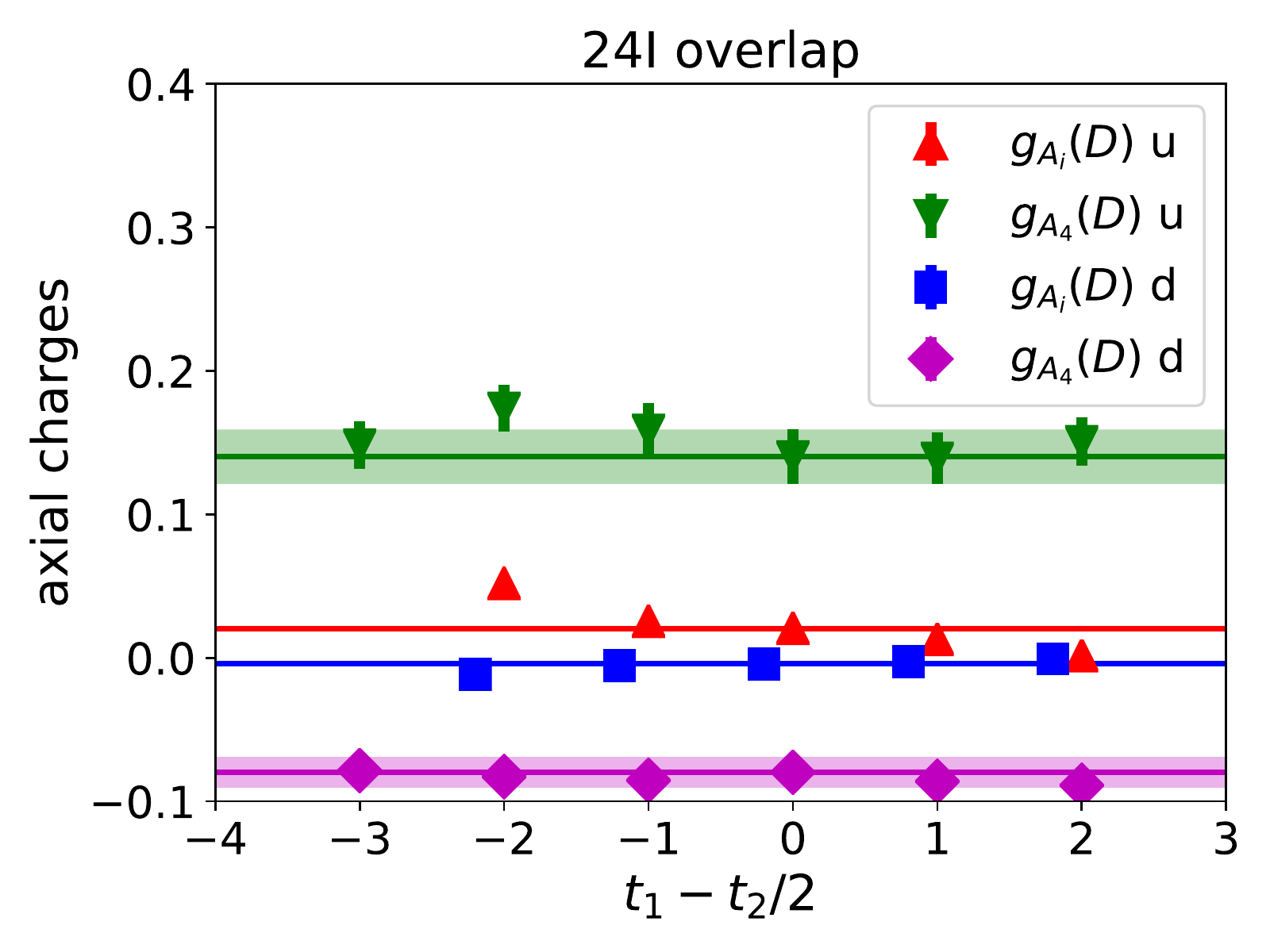}
\caption{The same as in Fig.~\ref{clover_test} for the overlap case, 24I. Pion mass is around $300$ MeV;  time separation is 8a.}
\label{overlap_singlet}
\end{figure}

The results of the clover test are listed in Fig.~\ref{clover_test}. 
For comparison, the separate $u$ and $d$ results of the overlap case with similar pion mass and source-sink separation 
are plotted in Fig.~\ref{overlap_singlet}. 
The following arguments can be made from these two figures:
\begin{itemize}
\item
The discrepancy exists in the clover case as well. We can see that in the overlap case, 
the gaps between the temporal and spatial components are around $0.1$ for both the $u$ quark and $d$ quark (left panel in Fig.~\ref{overlap_singlet}). 
In the clover case, although the results are noisier due to reduced statistics, the gaps are still consistent with $0.1$ for each of the two flavors. 
Especially for the $d$ quark case, the discrepancy is obvious (left panel in Fig.~\ref{clover_test}).
So we can conclude here that the difference between $g_{A_i}(P)$ and $g_{A_4}(P)$ is not an overlap fermion artifact 
or some mistake in the low-mode substitution sandwich code for the 3-point function contraction.
It may be a more common phenomenon for the point currents in different fermion actions.
\item
In the sandwich method for overlap, it is more expensive to evaluate $\overleftrightarrow{D}$, so we use $\overrightarrow{D}$ instead.
All the channels related to $\overrightarrow{D}_4$ show no conventional plateau,  e.g., the red dots in the right panel of Fig.~\ref{overlap_singlet} 
and the bottom left plot in Figs.~\ref{32I_gA_m3} and \ref{24I_gA_m3}.
Although a 2-state fit can deal with this very well (only needing the two coefficients $C_1$ and $C_2$ of the excited-state contamination to have opposite sign),
we use clover fermions to confirm the results. 
In the clover case, we directly use $\overleftrightarrow{D}$ and we can see (the red dots in the right panel of Fig.~\ref{clover_test}) a very flat plateau there, and the 
value is similar to that in the overlap case. This confirms that it is safe to use $\overrightarrow{D}$ for the overlap case if we conduct a 2-state fit in this channel.
\item
Figs.~\ref{clover_test} and \ref{overlap_singlet} show the results of $u$ quark and $d$ quark separately. Although the main topic of this work is
the isovector case, it is interesting to know whether our improvement scheme is still valid for the {individual flavors}, at least for the connected insertion part.
We can see that for the overlap case, the gaps between $g_{A_{i}}(D)$ and $g_{A_{4}}(D)$ for $u$ quark and $d$ quark are roughly $0.1$, 
which is the same as the gaps between $g_{A_{i}}$ and $g_{A_{4}}$ of the two flavors, 
meaning that the same factor $g$ also applies to the {individual flavor} within errors.
The situation of the clover case is similar; the improvement still works for each flavor.
So as mentioned before, when we focus on the CI part of the isosinglet axial charge for the quark spin calculation, the improvement of local currents is significant as well.
\end{itemize} 

\section{summary}
\label{summary}

As part of the effort to resolve the discrepancy of the isovector $g^3_A$ between lattice calculations and experiments, 
we employ dimension-4 operators to improve the local vector and axial vector currents and use these improved 
currents to calculate the nucleon axial charge on the lattice.
Numerical results show that for the vector cases, 
since $g_{V_i}(P)$ and $g_{V_4}(P)$ are consistent with each other within error bars, no improvement is needed in this channel.
For the axial vector cases, $g_{A_4}(P)$ is smaller than $g_{A_i}(P)$ by $\sim20\%$ and the difference is around $\sim7\sigma$, 
whereas the behaviors of the corresponding $g_{A_{i}}(D)$ and $g_{A_{4}}(D)$ are exactly the opposite,
leading to effective improvement. 
Using the equality of $g_{A_4}=g_{A_i}$ as a normalization condition, 
we find that the improved values of $g^3_A$ are increased by 3.4\% and 1.7\% for 24I and 32I
at the unitary point with final results of $g_A^3 = 1.22(4)(3)$ and $1.21(3)(3)$
for the 24I and 32I respectively. This is in the right direction for reducing the discrepancy between lattice calculations and experiments.

In addition to the control of excited-state contamination and the current improvement in this work,
continuum extrapolation, volume dependence and physical pion mass also need to be included to see if the $g^3_A$ discrepancy can be settled.
Furthermore, since the overlap fermion is a chiral fermion, we will be able to use the conserved axial current in future calculations,
which should give more reliable results from Lattice QCD.

\begin{acknowledgments}
We thank the RBC/UKQCD Collaborations for providing their DWF gauge configurations. 
This work is supported in part by the U.S. DOE Grant $\text{No.}$ $\text{DE-SC}0013065$. 
This research used resources of the Oak Ridge Leadership Computing Facility at the Oak Ridge National Laboratory, 
which is supported by the Office of Science of the U.S. Department of Energy under Contract No. DE-AC05-00OR22725. 
This work also used Stampede time under the Extreme Science and Engineering Discovery Environment (XSEDE)~\cite{6866038}, 
which is supported by National Science Foundation grant number ACI-1053575. 
We thank National Energy Research Scientific Computing Center (NERSC) for providing HPC resources that have contributed to the research results reported within this paper. 
We acknowledge the facilities of the USQCD Collaboration used for this research in part, which are funded by the Office of Science of the U.S. Department of Energy.
\end{acknowledgments}


\bibliographystyle{apsrev4-1}
\bibliography{tex}

\end{document}